\documentclass[11pt,preprint]{aastex}

\def\HI{\mbox{\ion{H}{1}}}
\def\HII{\mbox{\ion{H}{2}}}

\def\ie{{\it i.e.}}
\def\eg{{\it e.g.}}
\def\ltsima{$\; \buildrel < \over \sim \;$}
\def\simlt{\lower.5ex\hbox{\ltsima}}
\def\gtsima{$\; \buildrel > \over \sim \;$}
\def\simgt{\lower.5ex\hbox{\gtsima}}

\begin{document}

\title{Pre-reionization Fossils, Ultra-faint Dwarfs and the Missing Galactic Satellite Problem}
\author{Mia S. Bovill and Massimo Ricotti}
\affil{Department of Astronomy, University of Maryland, College Park,
  MD 20742}
\email{msbovill@astro.umd.edu,ricotti@astro.umd.edu}

\begin{abstract}
  We argue that, at least a fraction of the newly discovered
  population of ultra-faint dwarf spheroidal galaxies in the Local
  Group constitute the fossil relic of a once ubiquitous population of
  dwarf galaxies formed before reionization with circular velocities
  smaller than $v_{c}^{cr} \sim 20$~km/s. We present several arguments
  in support of this model. The number of luminous Milky Way
  satellites inferred from observations is larger than the estimated
  number of dark halos in the Galaxy that have, or had in the past, a
  circular velocity $>v_{c}^{cr}$, as predicted by the ``Via Lactea''
  simulation. This implies that some ultra-faint dwarfs are
  fossils. However, this argument is weakened by recent results from
  the ``Aquarius'' simulations showing that the number of Galactic
  dark matter satellites is 2.5 larger than previously
  believed. Secondly, the existence of a population of ultra-faint
  dwarfs was predicted by cosmological simulations in which star
  formation in the first minihalos is reduced -- but not suppressed --
  by radiative feedback.  Here, we show the statistical properties of
  the fossil galaxies in those simulations are consistent with
  observations of the new dwarf population and with the number and
  radial distribution of Milky Way satellites as a function of their
  luminosity.  Finally, the observed Galactocentric distribution of
  dwarfs is consistent with a fraction of dSphs being fossils. To
  make our case more compelling, future work should determine whether
  stellar chemical abundances of simulated ``fossils'' can reproduce
  observations and whether the tidal scenarios for the formation of
  Local Group dwarf spheroidals are equally consistent with all
  available observations.
\end{abstract}

\section{Introduction}

Hierarchical formation scenarios predict that most of the galactic
halos that formed before reionization (at $z=6-10$) had masses below
$10^8-10^9$ M$_{\odot}$.  Those that survived to the modern epoch, if
they were able to form stars, would constitute a sub-population of
dwarf satellites orbiting larger halos.  N-body simulations of cold
dark matter (CDM) predict a number of dark matter halos around the
Milky Way and M31 that is much greater than the number of observed
luminous satellites \citep{Klypinetal99, Mooreetal99}. This may
indicate a problem with the CDM paradigm or that feedback processes
are very efficient in suppressing star formation in the first small
mass halos, which remain mostly dark.  Recent observational and
theoretical advances require a re-visitation of the missing galactic
satellite problem.  In addition, cosmological simulations of the
formation of the first galaxies have shown most previously known dwarf
spheroidals (dSphs) to have properties consistent with the surviving
first galaxies, and predicted the existence of an undiscovered, lower
surface brightness population of dwarfs \citep{RicottiGnedin05}
(hereafter, RG05). The recent discovery of a population of ultra-faint
dwarfs confirms the aforementioned theoretical expectation and allows
us to test in great detail cosmological simulations of the first
galaxies, addressing important theoretical questions on feedback in
the early universe and on the nature of dark matter.

The formation of the first dwarf galaxies - before reionization - is
regulated by complex feedback effects that act on cosmological scales.
These self-regulation mechanisms have dramatic effects on the number
and luminosity of the first, small mass galaxies, and yet, are
unimportant for the formation of galaxies more massive than
$10^8$~M$_\odot$, that may form before and after reionization. Galaxy
formation in the high redshift universe is peculiar due to (i) the
lack of important coolants, such as carbon and oxygen, in a gas of
primordial composition and (ii) the small typical masses of the first
dark halos.  The gas in halos with circular velocity smaller than
$20$~km~s$^{-1}$ (roughly corresponding to a mass $M\simlt 10^8$
M$_{\odot}$ at the typical redshift of virialization) is heated to
temperatures $\simlt 10,000$ K during virialization.  At this
temperature, a gas of primordial composition is unable to cool and
initiate star formation unless it can form a sufficient amount of
H$_2$ ($x_{H_2} \simgt 10^{-4}$).  Although molecular hydrogen is
easily destroyed by far ultraviolet (FUV) radiation (negative
feedback), its formation can be promoted by hydrogen ionizing
radiation emitted by massive stars, through the formation of H$^-$
(positive feedback) \citep{HaimanReesLoeb96}.

It is difficult to determine the net effect of radiative feedback on
the global star formation history of the universe before the redshift
of reionization.  The effect of a dominant FUV background (at energies
between $11.34$~eV and $13.6$~eV) is to destroy H$_2$, the primary
coolant at the start of galaxy formation. The FUV radiation emitted by
the first few Population~III stars is sufficient to suppress or delay
galaxy formation in halos with circular velocities $v_{c}< v_c^{cr}
\sim 20$~km s$^{-1}$. According to this scenario, most halos with
masses $<10^8-10^9$ ~M$_\odot$ do not form stars and remain
dark. Therefore, the number of pre-reionization fossils in the Local
Group would be expected to be very small or zero.  However, this model
does not take into account the effect of ionizing radiation and
``positive feedback regions'' \citep{RicottiGnedinShull01, Ahnetal06,
  Whalenetal07}, that may have a dominant role in regulating galaxy
formation before reionization \citep{RicottiGnedinShull02a,
  RicottiGnedinShull02b}. Simulations including these processes show
that star formation in the first small mass halos is inefficient,
mainly due to winds produced by internal UV sources.  This produces
galaxies that are extremely faint and have very low surface
brightnesses.  However, according to the results of our simulations, a
large number of these ultra-faint dwarfs (a few hundred galaxies per
co-moving Mpc$^3$) form before reionization at $z \sim 7-10$. Hence,
according to this model, the Local Volume and the Local Group should
contain hundreds of ultra-faint dwarf galaxies.

The small masses of the first minihalos have two other implications.
First, the ionizing radiation emitted by massive stars can blow out
most of the gas before SN-driven winds become important, further
reducing star formation rates \citep{RicottiGnedinShull08}.  Second,
the increase in temperature of the intergalactic medium (IGM) to
$10,000-20,000$ K due to \HI~reionization, prevents the gas from
condensing into newly virialized halos with circular velocities
smaller than $10-20$~km~s$^{-1}$, with a critical value that depends
on redshift \citep{Gnedin-filteringmass00, OkamotoGT08}. It follows,
that dwarf galaxies with $v_c < 10-20$~km~s$^{-1}$ stop forming stars
after reionization. However, the value $v_c^{cr}\sim 20$~km~s$^{-1}$
that we use to define a fossil is primarily motivated by the
fundamental differences discussed above in cooling and feedback
processes that regulate star formation in the early Universe and is
not the critical value for suppression of star formation due to
reionization.  Indeed, \cite{Ricotti:08} argues that fossils dwarfs
can have a late phase of gas accretion and star formation well after
reionization, at redshift $z<1-2$. Thus, a complete suppression of
star formation after reionization (about 12~Gyr ago) is not a defining
property of a fossil dwarf.

Data on the velocity dispersion of the stars, $\sigma$, the only
observational measure of halo mass, shows that a typical dSph has
$\sigma <20$~km~s$^{-1}$. However, the circular velocity of the dark
halo can be much larger than $\sigma$ if typical radius of the stellar
spheroid is much smaller than the dark halo radius. For a
pre-reionization fossil, according to simulations in RG05, on average
we measure $\sigma/v_c \sim 0.5$ at formation (see Fig.~2 in RG05).

RG05 compared the statistical properties of simulated pre-reionization
galaxies to observations of Local Group dwarfs available in 2004-2005.
Based on similarities between observed dSphs and simulated galaxies
formed before reionization, they argued that many dSphs may be
``fossils'' of the first galaxies.  RG05 also predicted the existence
of a yet undetected population of ultra-faint dSph galaxies.
\citep{GnedinKravtsov06} (hereafter, GK06) used results from RG05 to
predict the radial distribution of fossils around the Milky Way and
their Galactocentric luminosity function. They found a good agreement
of simulations with observations for the most luminous fossils but,
again, a lack of observed ultra-faint fossils, mainly in the outer
Milky Way halo.  An ultra-faint population of dSphs has now been found
in the Local Group.  As we will show in the present work, these new
dwarfs have properties in agreement with our simulations of
pre-reionization fossils. This discovery is certainly one of the most
exciting developments in understanding galaxy formation in the early
universe and has drawn renewed attention to ``near field cosmology''
as a tool to understand galaxy formation.  The new galaxies have been
discovered by data mining the SDSS and surveys of the halo around M31,
resulting in the discovery of 12 new ultra-faint Milky Way satellites
\citep{Belokurovetal06a, Belokurovetal07, Irwinetal07, Walshetal07,
  Willmanetal05ApJ, Willmanetal05AJ, Zuckeretal06a, Zuckeretal06b, Gehaetal08} and
six new companions for M31 \citep{Ibataetal07, Majewskietal07,
  Martinetal06}. 

Here, we also argue that solely based on the observed number of
Milky-Way satellites, at least a few of them must be a
pre-reionization fossil, that means: it must have formed before
reionization in a halo with circular velocity $<20$ km/s. However,
some ultra-faint dwarfs may not be pre-reionization fossils. As we
write, several works have been published that seem to show that the
observed properties of the ultra-faint dwarf population can also be
explained in the context of the tidal scenario, that assumes that
these galaxies formed after reionization in halos that were much more
massive and had different properties at the time of formation
\citep{PenarrubiaNM08}.  It is intriguing that both the tidal scenario
and the pre-reionization fossil scenario are able to produce a
population of ultra-faint dwarfs that follow very similar statistical
trends in terms of size, surface brightness, mass to light ration and
metallicity-luminosity relation. The jury is still out.

This paper is laid out as follows. In \S~\ref{sec:data}, we collect
published data on the new dwarf population and, after correcting for
completeness of the surveys, we estimate the total number of Local
Group satellites (which increases from 32 to about 100). Using the
results of published N-body simulations, we compare the observed
number of luminous satellites to the estimated number of dark
satellites that have or had in the past a circular velocity
$>v_{c}^{cr}$, using the results of published N-body simulations,
concluding that some ultra-faint dwarfs must be pre-reionization
fossils.  In \S~\ref{sec:comp} we show that the properties of the new
Milky Way and M31 dwarfs are in remarkable agreement with the
theoretical data on the ``fossils'' from RG05 and with their
Galactocentric distribution around the Milky Way calculated in GK06.
In \S~\ref{sec:disc} we discuss the implications of the new dwarfs on
the formation of the first galaxies and the missing galactic satellite
problem.

\section{Data and Completeness Corrections}\label{sec:data}

In Table~1, we summarize the observed properties of the new dwarfs.
The new Milky Way satellites were discovered using SDSS Data Release 4
and 5 \citep{Adelman-McCarthyetal06,Adelman-McCarthyetal07}.  When
multiple references are available for a dwarf property, we defer to
the measurement with the smallest error bars.  Excepting Bootes I and
II, Canes Venatici I and Leo T, where central surface brightness
measurements were available, the average surface brightness inside the
half light radius, $r_{1/2}$, was used: $\Sigma_{V} = L_{V}/(2 \pi
r_{1/2}^{2})$.  

Recent surveys of M31 \citep{Martinetal06,Ibataetal07} have covered
approximately a quarter of the space around the M31 spiral.  The
survey have found 6 new M31 satellites. If we make a simple correction
for the covered are of the survey we find that, including the new
dwarfs, the estimated number of M31 satellites increases from $9$ to
$33 \pm 10$.  Two new M31 satellites, And XII and And XIV, have
velocities near or above their host's escape velocity
\citep{Chapmanetal07,Majewskietal07}.  Both galaxies are classified as
dwarf spheroidals and show a lack of \HI~gas, and both are likely on
their first approach towards a massive halo.  Kinematic data is not
yet available on these two dwarfs to determine whether their circular
velocities are below the $20$~km~s$^{-1}$ threshold, however their
currently known properties meet the RG05 criteria for fossils.

In estimating the completeness correction for the number of the Milky
Way dwarfs, one should also account for selection effects from the
limiting surface brightness sensitivity of the Sloan of $\sim$30 ${\rm
  mag~arcsec}^{-2}$ \citep{Koposovetal07}. The sensitivity limit is
shown as a solid line in Figure~\ref{Kor}.  Identification
of new satellites depends on the visibility of the horizontal branch
in the color-magnitude diagram, which, for the typical luminosity of
the new faint dwarfs ($M_V \approx -4$) drops below SDSS detection
limits at Galactocentric distances beyond $\sim 200-250$~kpc
\citep{Koposovetal07}.  Of the new Milky Way dwarfs, only Leo T is
well beyond this distance threshold and nine of the eleven new Milky
Way satellites are within $200$~kpc. We make the most conservative
estimate, by assuming that we have a complete sample of dwarfs within
$200$~kpc.  Additional selection bias for the new dwarfs comes
primarily from the limits of the SDSS coverage on the sky.  To account
for this, we apply the zero-th order correction of multiplying the
number of new dwarfs by 5.15 \citep{Tollerudetal08}. This correction
assumes an isotropic distribution of satellites when observed from the
Galactic center. With this simple assumptions we estimate that the
number of Milky Way satellites with Galactocentric distance $<200$~kpc
is about $85 \pm 14$, including the 29 previously know satellites. The
error estimate is due to shot noise.

However, bright satellites of the Milky Way are distributed very
anisotropically \citep{Kroupa:05, Zentner:05}, so the assumption of
isotropy may not be a good one.  In addition, the luminous satellites
can be radially biased, so the abundance of the faintest satellites
within 50~kpc may not be easily corrected to larger distances without
prior knowledge of this bias. And, of course, satellites of different
luminosity and surface brightness will have different completeness
limits.  These selection biases have been considered in detail in a
recent paper by \cite{Tollerudetal08}.  This study finds that there
may be between 300 to 600 luminous satellites within the virial
radius of the Milky Way.  Their estimate for the number of luminous
satellites within a Galactocentric distance of about $200$~kpc is 120,
that is slightly larger than our simple (and more conservative)
estimate.

\subsection{Number of non-fossil satellites in the Milky Way}\label{ssec:count}

In this section, we use the results of published N-body simulations to
estimate the number of dark halos in the Milky Way that have, or had, a circular velocity $v_c>20$~km/s. By definition, dwarf
galaxies formed in these dark halos are not pre-reionization fossils.
If we find that the number of observed Milky Way satellites exceeds
the estimated number of these massive halos we must conclude that at
least a fraction of the observed Milky Way satellites are
pre-reionization fossils. GK06 have estimated that pre-reionization
fossils may constitute about $1/3$ of Milky Way dwarfs, based on
detailed comparisons between predicted and observed Galactocentric
distributions of dwarf satellites.

It is clear that if we simply count the number of dark halos within
the Milky Way virial radius with $v_c \simgt 20$~km/s, their number is
much smaller than the current number of observed luminous
satellites. However, a significant fraction of dark halos that today
have $v_c < 20$~km/s were once more massive, due to tidal tripping
\citep{KravtsovGnedinKlypin04} . If the stars in these halos survive
tidal stripping for as long as the dark matter, they may indeed
account for a fraction or all of the newly discovered ultra-faint
dwarfs. \cite{KravtsovGnedinKlypin04} favor the idea that tidal
stripping of the dark matter halo does not affect the stellar
properties of the dwarf galaxy. Thus, this model is qualitatively
similar to our model for pre-reionization fossils, save a rescaling of
the mass of the dark halos hosting the dwarfs.

However, it is also possible that tidally stripped halos loose their
stars more rapidly than they loose their dark matter. Thus, they may
quickly transform from luminous to dark halos. Such a behavior has
been found by \cite{PenarrubiaNM08}. According to this scenario,
tidally stripped dark halos may not account for the observed
ultra-faint population. To summarize, if the number of dark halos that
have or had in the past $v_c \simgt 20$~km/s is smaller than the
number of luminous Milky Way satellites we may conclude that some dwarfs
are fossils. Vice versa, if the number is larger, we cannot make any
conclusive statement about the origin of Milky Way satellites.

High resolution N-body simulations of the Milky Way system give the
number of dark halos in the Milky Way as a function of their circular
velocity $v_{c}$ at $z=0$.  The ``Via Lactea'' simulation by
\cite{Diemandetal07a} finds:
\begin{equation}
N_{dm}(>v_{c})= N_{dm, 20} \left({v_{c} \over 20 km/s}\right)^{-\alpha},
\end{equation}
with $N_{dm, 20} \approx 27.7$ and $\alpha \approx 3$. However, a
recent work by \cite{Springeletal08} (the Aquarius simulations) finds
a factor 2.5 more satellites at any given $v_{c}$, \ie, $N_{dm, 20}
\approx 69$ and $\alpha \approx 3.15$. Although the Aquarius
simulations have higher resolution than the Via Lactea simulation, the
large disagreement between the two works is due to a systematic
difference, possibly related to the creation of the initial
conditions, and it is not due to the improved resolution. Although the
Aquarius simulation is likely correct, we will provide predictions for
both simulations (we have become aware of the Aquarium simulation
results, that are not published yet, after this work had been mostly
completed.)

To determine the importance of tidal mass loss for satellites around
the Milky Way we use results from \cite{KravtsovGnedinKlypin04}.
Figure~5 in \cite{KravtsovGnedinKlypin04} gives the fraction of halos,
$f(v_{c})$, that presently have circular velocity $v_c$, but some time
in the past had a circular velocity $\le v_{c}^{max}=20$~km/s, where
as $v_{c}^{max} \equiv \max(v_{c}(t))$.  We approximate the
\cite{KravtsovGnedinKlypin04} results for $f(v_c)$ with the power law
$f(v_c) \approx (v_c/20$~km~s$^{-1})^{\beta}$, with $\beta \approx
3.7$. We then calculate the number of dark halos $N_{dm}(v_{c}^{max} > 20$~km~s$^{-1})$
analytically:
\begin{eqnarray*}
N_{dm}(v_{c}^{max} > 20~{\rm km~s}^{-1}) &=& N_{dm}(v_{c}(z=0) > 20~{\rm km~s}^{-1}) +
\int^{20~km/s}_{v_{c, min}} dv \frac{dN}{dv} f(v)\\ 
&=&N_{dm, 20}\left[{1+{\alpha \over \beta-\alpha}(1-x_{min}^{\beta-\alpha})}\right] \approx 2.64 N_{dm, 20}
\end{eqnarray*}
where $x_{min}=v_{min}/20~{\rm km/s}$, and $v_{min} \simgt <\sigma_*>
\approx 10$~km~s$^{-1}$ roughly equals the mean observed velocity
dispersion of the stars, $<\sigma_*>$, of ultra-faint dwarf satellites. The
rationale for integrating to $v_{min}$ is that observed satellites are
dark matter dominated and cannot be hosted in dark halos that have
$v_{c} < \sigma_*$, unless $\sigma_*$ is not a tracer for the dark
matter content of the halo (\eg, due to tidal heating).

Using the above equation, we find $73 \pm 16$ and $182 \pm 40$ halos
with $N_{dm}(v_{c}^{max} > 20$~km~s$^{-1})$ within $R_{vir}$, for the
Via Lactea and Aquarius simulations respectively. Both these numbers
are smaller than the $300-1000$ luminous Milky Way satellites estimated by \cite{Tollerudetal08}. Taken at face value,
these numbers indicate that a fraction of Milky Way satellites are
true pre-reionization fossils. However, the number of luminous
satellites that exist within the Milky Way is highly uncertain beyond
a distance from the Galactic center of $200$~kpc.

Based on Figures~11 and 12 in \cite{Springeletal08} and Figure~5 in
\cite{Tollerudetal08}, we estimate that roughly half of the Milky Way
satellites (within the virial radius $R_{vir}$) are
$<200$~kpc from the Galactic center.  Therefore, within $200$~kpc we
estimate $36 \pm 8$ and $91 \pm 20$ dark halos with $N_{dm}(v_{c}^{max} >
20$~km~s$^{-1})$ for the Via Lactea and Aquarius simulations
respectively. These numbers can be compared to our estimated number of
luminous satellites with $d<200$~kpc ($85 \pm 14$ satellites) and to
the estimate by \cite{Tollerudetal08} ($120$ satellites). Using the
Via Lactea simulation, we still find that some dwarfs are true
pre-reionization fossils but the argument is weak if we use the
Aquarius simulation results.  In Table~\ref{tab:count} we summarize
the counts for dark matter and luminous satellites discussed in this
section.

Although there is considerable uncertainty in our estimates, it can be
safely concluded that, using the results of the Via Lactea simulation,
at least a fraction of Milky Way dwarfs are fossils. However, this
argument does not hold anymore or is weakened due to recent results
(not yet published in a refereed journal) from the Aquarius
simulations, showing a factor 2.5 increase for the number of Milky Way
dark matter satellites in any mass range.

\begin{deluxetable}{l|cc|ccccc}
\tablecaption{Number of observed satellites versus number of dark halos with $v_{c}^{max}(t)>20$~km/s (\ie, non pre-reionization fossils) for the Milky Way}\label{tab:count}
\tablewidth{0pt}
\tablehead{
\colhead{} &
\multicolumn{2}{|c|}{number of luminous dwarfs} &
\multicolumn{4}{|c}{number of dark halos $N_{dm}$ with $v_{c}^{max}>20$~km/s}\\
\colhead{distance from} & \multicolumn{2}{|c|}{} & 
\multicolumn{2}{|c}{Via Lactea sim.} &
\multicolumn{2}{c}{Aquarius sim.}\\
\colhead{Galactic center} & \colhead{this work} &
\colhead{Tollerud et al.} &
{today} & \colhead{any time} &
{today} & \colhead{any time}}

\startdata
$d<200$~kpc & $85 \pm 14$ & 120 & 14 & $36 \pm 8$ & 34 & $91 \pm 20$ \\
$d<R_{vir}\sim 400$~kpc & \nodata & from 300 to 600 & 28 & $73 \pm 16$ & 69 & $182 \pm 40$ \\
\enddata
\end{deluxetable}

\subsection{Peculiar ultra-faint dwarfs}

Almost all newly discovered dwarfs are dSphs with a dominant old
population of stars and virtually no gas, which makes them viable
candidates for being pre-reionization fossils. However, there are
two notable exceptions that we discuss below.

\subsubsection{Leo~T}

With the gas and young stars of a typical dIrr and the radius,
magnitude, mass and metallicity of a dSph \citep{Irwinetal07,
  SimonGeha07}, Leo~T presents a puzzle.  Leo~T has a stellar velocity
dispersion of $\sigma_{Leo T} = 7.5 \pm 1.6 {\rm km~s}^{-1}$
\citep{SimonGeha07}, or an estimated dynamical mass of
$10^7$~M$_\odot$ within the stellar spheroid (although its total halo
mass may be much larger).  Leo~T shows no sign of recent tidal
disruption by either the Milky Way or M31 \citep{deJongetal08} and is
located in the outskirts of the Milky Way at a Galactocentric distance
of $400$~kpc. Leo~T photometric properties are identical to those of
pre-reionization fossils. However, if we assume that Leo~T is a
pre-reionization fossil, it is not expected to retain significant gas
or form stars after reionization.

How did Leo~T keep its \HI~and how did its $<9$~Gyr old stellar
population form?  Work by \cite{Stinsonetal07} suggests cyclic heating
and re-cooling of gas in dwarf halos can produce episodic bursts of
star formation separated by periods of inactivity.  The lowest
simulated dwarf to form stars has $\sigma = 7.4 {\rm km~s}^{-1}$,
similar to Leo~T. However, the increasing of the IGM Jeans mass after
reionization should prevent gas from condensing back into halos with
circular velocity $v_c <20$~km~s$^{-1}$. Another proposal (Ricotti
2008) is that, although the mass of Leo~T at formation was $\sim
10^7-10^8$~M$_\odot$ (\ie, a fossil), as indicated by its stellar
velocity dispersion, its present dark matter mass and the halo
concentration has increased after virialization by roughly a factor of
$10/(1+z)$. This is expected if Leo T has evolved in isolation after
virialization, as it seems to be indicated by its large distance from
the Milky Way.  In this scenario, Leo~T stopped forming stars after
reionization, but it was able to start accreting gas again from the
IGM very recently (at $z \simlt 1-2$). This can explain the $<9$~Gyr
old stellar population and the similarity of Leo~T to the other
pre-reionization fossils.

\subsubsection{Willman I}

Assuming virial equilibrium Wilman~I has a dynamical mass within the
largest stellar orbit ($r \sim 100$~pc) of $5 \times 10^{5} M_{\odot}$
and a M/L of $\sim 470$, similar to other ultra-faint dwarfs. However,
Willman~I has a distance from the Galactic center, luminosity, surface
brightness and size typical of a galactic globular clusters
\citep{Willmanetal05AJ}.  The combination of dark matter domination
with a small core radius, and a mass an order of magnitude smaller
than the other SDSS dwarfs suggests Willman~I exists in the region
between dwarf spheroidals and globular clusters.  For our discussion
of the missing galactic satellites, we treat Willman~I as a dwarf
because of its large dark matter fraction. However, its properties do
not agree with being a ``fossil'' of the first galaxies (see
Figure~\ref{Kor} in the next section).

\section{Comparison with Theory}\label{sec:comp}

In this section, we compare the properties of the new dwarf galaxies
discovered in the Local Group to the theoretical predictions of
simulations of primordial galaxies formed before reionization. The
argument that justifies this comparison is as follows.

After reionization, due to IGM reheating, the formation of galaxies
smaller than $20$~km~s$^{-1}$ is inhibited because the thermal
pressure of the IGM becomes larger than the halo gravitational
potentials \citep[\eg,][]{Gnedin-filteringmass00}. Galaxies formed
before reionization stop forming stars due to the progressive
photo-evaporation of their interstellar medium by the ionizing
radiation background (although most of the ISM was already lost due to
UV driven galactic winds and SN explosions). Hence, if
pre-reionization dwarfs do not grow above $v_c=20$~km~s$^{-1}$ by
mergers, their stellar population evolves passively as calculated by
stellar evolution models such as Starburst99 \citep{Leithereretal99}.
We define such galaxies as pre-reionization ``fossils''.

Clearly, we do not expect two perfectly distinct populations of fossils
with $v_c<20$~km~s$^{-1}$ and non-fossils with $v_c\ge
20$~km~s$^{-1}$, but a gradual transition of properties from one
population to the other.  Some fossils may become more massive than
$v_c \sim 20$~km~s$^{-1}$ after reionization, accrete gas from the
IGM, and form a younger stellar population. If the dark halo circular
velocity remains close to $20$~km~s$^{-1}$ the young stellar
population is likely to be small with respect to the old one. We call
these galaxies ``polluted fossils'' because they have the same basic
properties of ``fossils'' with a sub-dominant young stellar population
(see RG05).

Vice versa, some non-fossil galaxies with $v_c>20$~km~s$^{-1}$ may
loose a substantial fraction of their mass due to tidal interactions.
If they survive the interaction, their properties, such as surface
brightness and half light radius, may be either modified or stay the
same.  \cite{KravtsovGnedinKlypin04} estimate that 10$\%$ of Milky Way
dark matter satellites were at least ten times more massive at their
formation than they are today and more were a few times more massive
than they are today. Although their simulation does not include stars,
they favor the idea that the stellar properties of these halos would
be unchanged. Conversely, a recent work of \cite{PenarrubiaNM08} looks
at tidal stripping of dark matter and stars, achieving some success in
reproducing the observed properties of ultra-faint dwarfs assuming
that they are tidally stripped dIrrs.  Using our simulations, we
cannot make predictions of the internal properties of non-fossil
dwarfs, which are too massive to be present in significant numbers in
the small volume of our simulation.  However, using our simulation
data (in RG05), GK06 finds that about one third of Milky Way satellites
may be fossils based on comparisons between observed and simulated
Galactocentric distribution of the satellites.

\subsection{Description of the simulation}

Given a cosmological model, simulating the formation of the first
stars is a relatively well defined initial condition problem. However,
these simple initial conditions become unrealistic as soon as a few
stars form within a volume of several thousands of co-moving
Mpc$^{3}$. The physics becomes complex as competing feedback effects
determine the fate of the first galaxies: radiative feedback regulates
the formation and destruction of H$_2$ in the intergalactic medium and
in protogalaxies.  The background in the Lyman-Werner bands (between
$11.3$~eV and $13.6$~eV) dissociates H$_2$ through the two-step
Solomon process. A few Population~III stars per Mpc$^3$ can suppress
or delay the formation of the first small mass galaxies, reducing
drastically their number per unit co-moving volume
\citep{Haimanetal00,Ciardietal00,Machaceketal00}.

Ionizing radiation may enhance the production of H$_2$ (we refer to
this as ``positive feedback'') by creating free electrons and
promoting the formation of H$^-$, the main catalyst for the formation
of H$_2$ \citep*{ShapiroKang87, HaimanReesLoeb96,
  RicottiGnedinShull01, AlvarezBrommShapiro06, Ciardietal06}.
\cite{RicottiGnedinShull01} studied the structure of \HII~regions in
the early universe in a gas of primordial composition.  It was found
that shells of H$_2$ can be continuously created in precursors around
the Str\"omgren spheres produced by ionizing sources and, for a
bursting mode of star formation, inside recombining \HII~regions.  This is because the catalyst H$^-$, and hence H$_2$, is formed in regions where the gas
ionization fraction is about $50\%$. The thickness of these regions
depends on the density of the gas and the spectrum of the ionizing
radiation.  This local ``positive feedback'' may be the dominant
feedback, but it is difficult to incorporate into cosmological
simulations because the implementation of spatially inhomogeneous,
time-dependent radiative transfer is computationally expensive.

The simulation analyzed in the present paper differs from other
studies because it includes ``positive feedback'' from ionizing
radiation self-consistently \citep{RicottiGnedinShull02a,
  RicottiGnedinShull02b, RicottiGnedinShull08}(hereafter, RGS02a,b,
RGS08). The simulation have a space resolution of 156 pc~h$^{-1}$
comoving (about $15$~pc physical at $z=10$) and a mass resolution of
$4.93\times 10^3$~M$_\odot$~h$^{-1}$ for dark matter and $\approx
657$~M$_\odot$~h$^{-1}$ for the baryons. There are more than $10^5$
stellar particles in our simulations at $z \sim 10$. The stellar
masses are always smaller than the baryon mass resolution but can vary
from $\sim 0.6$~M$_\odot$~h$^{-1}$ to $600$~M$_\odot$~h$^{-1}$ with a
mean of $6$~M$_\odot$~h$^{-1}$. Stellar particles do not represent
real stars. In addition to primordial chemistry and 3D radiative
transfer, the simulations include a recipe for star formation, metal
production by SNe and metal cooling (see RGS02a for details). The code
also includes mechanical feedback by SN explosions. However, we found
that for a Salpeter IMF, the effect of SNe is not dominant when
compared to feedback produced by ionizing radiation from massive stars
\citep{RicottiGnedinShull08}. The effect of SN explosion is somewhat
model dependent and uncertain because it is treated using a sub-grid
recipe. Hence, the simulation analyzed in this work includes metal
pollution but not mechanical feedback by SNe.

RGS02b have shown that the main negative feedback thought to suppress
the formation of the first galaxies (H$_2$ photo-dissociation) is not
dominant, contrary to the results of previous studies that did not
include ``positive feedback'' from ionizing radiation.  Feedback by
ionizing radiation plays the key role, inducing a bursting star
formation mode in the first dwarf galaxies. Galactic outflows,
produced by UV photo-heating from massive stars, and H$_2$
formation/photo-dissociation induce the bursting star formation mode
that acts as the catalyst for H$_2$ re-formation inside relic
(recombining) \HII~regions and in the ``precursors'' of cosmological
Str\"omgren spheres ({\it i.e.}, positive feedback regions).  As a
result, star formation in the first galaxies is self-regulated on a
cosmological scale - it is significantly reduced by feedback but it is
not completely suppressed, even in small mass halos with $v_c \sim
5-10$~km~s$^{-1}$. Due to the feedback-induced bursting mode of star
formation in pre-reionization dwarfs, the cosmological \HII~regions
that they produce remain confined in size and never reach the overlap
phase that defines the epoch of reionization.

The simulation data shown in this paper is the higher resolution run
in RGS02b, evolved further to redshift $z=8.0$ after the introduction
of a bright source of ionizing radiation that completes reionization
at $z \sim 9$ (see RG05 for details). The need for introducing a
bright ionization source is dictated by the small volume of the
simulation ($1.5^3$ Mpc$^3$); otherwise the volume would be reionized
too late. The \HI~ ionizing source removes all the remaining gas from
halos with $v_c<20$~km~s$^{-1}$ and shuts down star formation.

\subsection{Statistical properties of simulated ``fossils'' vs observations}\label{sec:prop}

\begin{figure}
\plotone{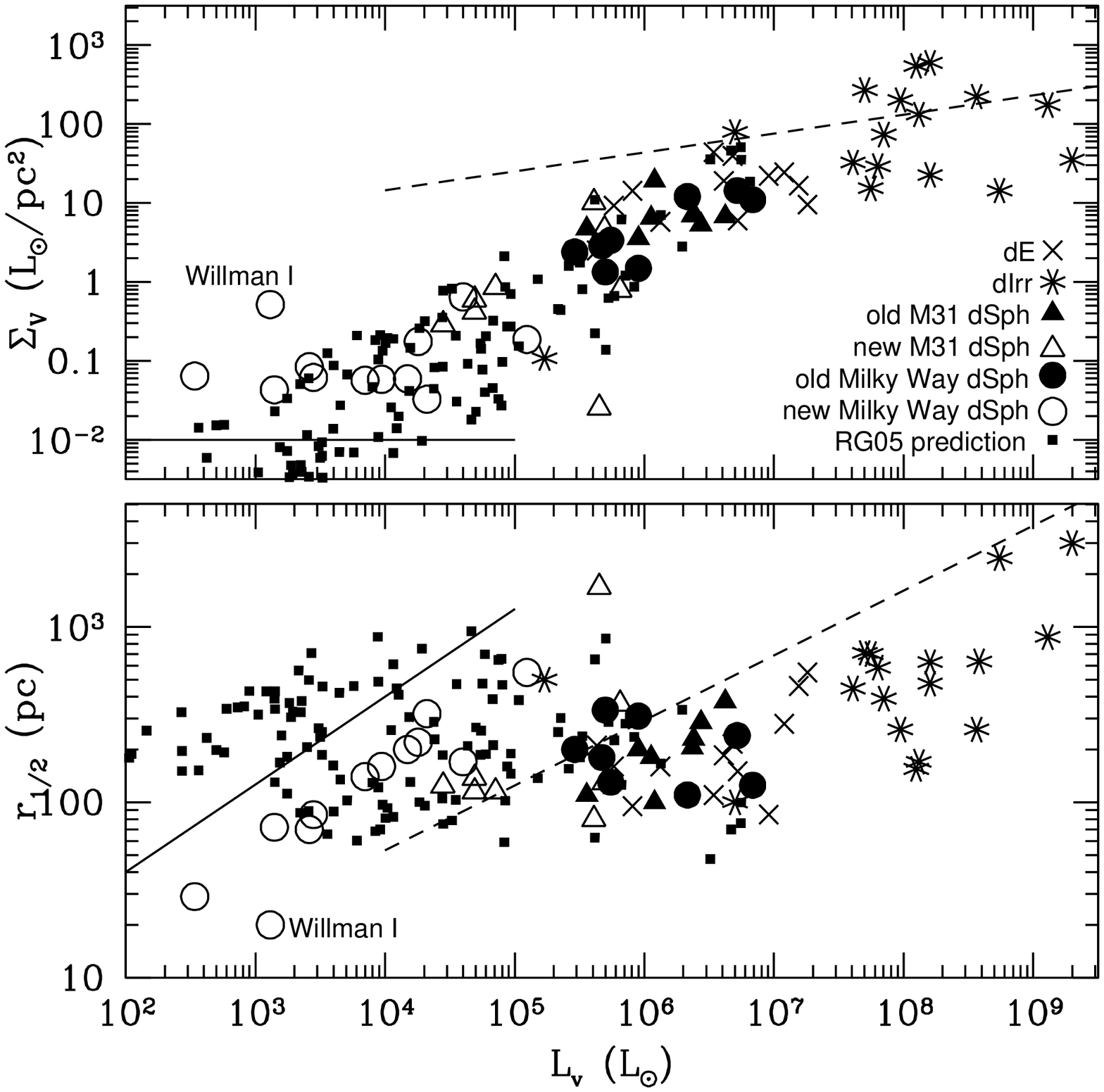}
\caption{An extension of Figure 7 from RG05 to include new dwarfs in
  the SDSS
  \citep{Belokurovetal07,Belokurovetal06a,Irwinetal07,Willmanetal05ApJ,Willmanetal05AJ,Walshetal07,Zuckeretal06a,Zuckeretal06b,Gehaetal08}
  and recent surveys of Andromeda
  \citep{Ibataetal07,Majewskietal07,Martinetal06}.  Surface brightness
  and core radius are plotted vs. V-band luminosities.  Small filled
  squares are the RG05 predictions, asterisks are known survivors,
  crosses are known polluted fossils, closed circles are the
  previously known dSph around the Milky Way, closed triangles are
  previously known dSph around M31, and open circles and triangles are
  new dSph around the Milky Way and M31 respectively. The solid lines
  show the SLOAN surface brightness limits and the dashed lines show
  the scaling relationships for more luminous Sc-Im galaxies
  ($10^8L_\odot \simlt L_B \simlt 10^{11}L_\odot$) derived by
  \cite{Kormendy:04}.}
   \label{Kor}
\end{figure}

Here we compare the RG05 predictions for the fossils of primordial
galaxies to the observed properties (see Table~\ref{tab:one}) of the
new Milky Way and M31 dwarfs. The symbols and lines in
Figs.~\ref{Kor}-\ref{ZS} have the following meanings.  All known Milky
Way dSphs are shown by circles; Andromeda's dSphs satellites are shown
by triangles; simulated fossils are shown by the small solid squares.
The solid and open symbols refer to previously known and new dSphs,
respectively. The transition between fossils and non-fossil galaxies
is gradual. In order to illustrate the different statistical trends of
``non-fossil'' galaxies we show dwarf irregulars (dIrrs) with
asterisks and the dwarf ellipticals (dE) as crosses, and we show the
statistical trends for more luminous galaxies as thick dashed lines on
the right side of each panel.

Figure~\ref{Kor} shows how the surface brightness (top panel) and half
light radius (bottom panel) of all known Milky Way and Andromeda
satellites as a function of V-band luminosity compares to the simulated
fossils.  The surface brightness limit of the SDSS is shown by the
thin solid lines in both panels of the figure.  The new dwarfs agree
with the predictions up to this threshold, suggesting the possible
existence of an undetected population of dwarfs with $\Sigma_{V}$
below the SDSS sensitivity limit.  The new M31 satellites have
properties similar to their previously known Milky Way counterparts
(\eg, Ursa Minor and Draco).  Given the similar host masses and
environments, is reasonable to assume a similar formation history for
the halos of M31 and the Milky Way.  This suggests the existence of an
undiscovered population dwarfs orbiting M31 equivalent to the new SDSS
dwarfs.


\subsubsection{Extreme Mass-to-Light ratios}\label{sec:ML}

\begin{figure}
\plotone{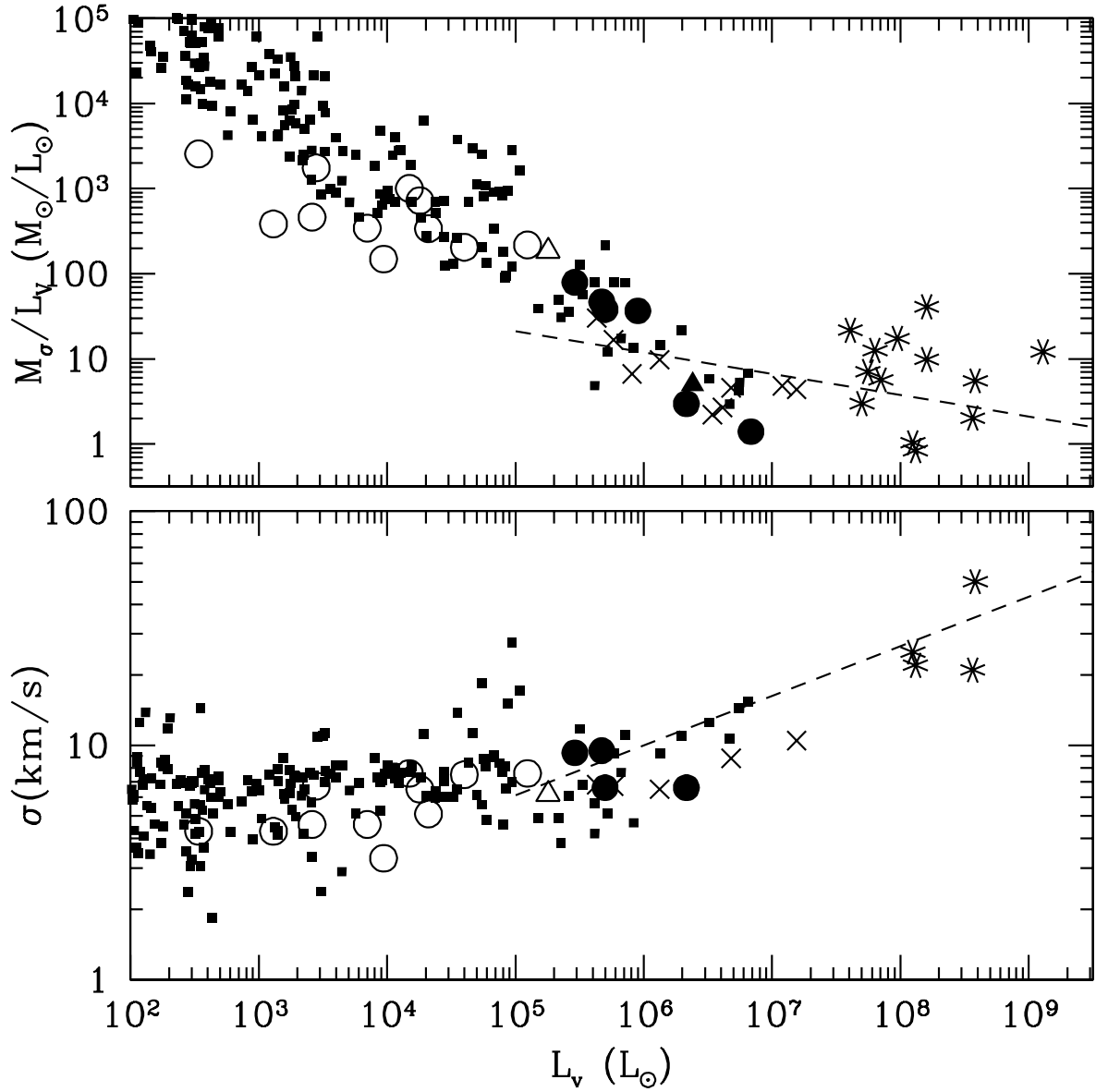}
\caption{Mass-to-light ratio and velocity dispersion of a subset of
  the new dwarfs \citep{Martinetal07,SimonGeha07} plotted with values
  for previously known dwarfs and RG05 predictions. The dashed lines
  show the scaling relationships for more luminous Sc-Im galaxies
  ($10^8L_\odot \simlt L_B \simlt 10^{11}L_\odot$) derived by
  \cite{Kormendy:04}.}
   \label{SL}
\end{figure}

The large mass outflows due to photo-heating by massive stars and the
consequent suppression of star formation after an initial burst, make
reionization fossils among the most dark matter dominated objects in
the universe, with predicted M/L ratios as high as $10^4$ and $L_V \sim
10^3-10^4$~L$_{\odot}$.

Figure~\ref{SL} shows the velocity dispersion (bottom panel) and
mass-to-light ratios, $M/L_V$ (top panel), as a function of V-band
luminosity of the new and old dwarfs from observations in comparison
to simulated fossils.  The symbols are the same as in the previous
figures.  While mass data is available for all the previously known
dwarfs, we found no published $\sigma$ values for 9 dIrr, 4 dE and 3
dSph (Antila, Phoenix and SagDIG) \citep{Mateo98, Strigarietal08}.  We
observe a good agreement between the statistical properties of the new
dwarf galaxies and the RG05 predictions for the fossils, although
simulated dwarfs show a slightly larger mass-to-light ratios than
observed ones at the low luminosity end, $L_V <10^4$~L$_\odot$.
Theoretical and observed dynamical masses are calculated the same way,
from the velocity dispersions of stars (\ie, $M=2r_{1/2}\sigma^2/G$),
and do not necessarily reflect the total mass of the dark halo at
virialization. Indeed, the simulation provides some insight on why the
observed value of the dynamical mass, $M \sim (1 \pm 5) \times
10^7$~M$_\odot$, remains relative constant as a function of $L_V$.
Simulations show that in pre-reionization dwarfs, the ratio of the
radius of the stellar spheroid to the virial radius of the dark halo
decreases with increasing dark halo mass.  The lowest mass dwarfs have
stellar speroids comparable in size to their virial radii (see RGS08).
As the halo mass and virial radius increases, the stellar spheroid
becomes increasingly concentrated in the deepest part of the potential
well, thus the ratio of the dynamical mass within the largest stellar
orbits to total dark matter mass is reduced. This effect maintains the
value of the dynamical mass within the stellar spheroid (measured by
the velocity dispersion of the stars) remarkably constant even though
the total mass of the halo increases.

If dwarfs are undergoing tidal disruption (\eg, Ursa Major II), the
velocity dispersions could be artificially inflated.  However, the
agreement with theory is rather good for all the new ultra-faint
dwarfs.  The data on the lowest luminosity dwarfs in our simulations
are the least reliable, because they are very close to the resolution
limits of the simulation (we resolve halos of about $10^5$~M$_\odot$
with 100 particles). Let us assume we can trust the simulation and the
observational data, even for the lowest luminosity dwarfs and the
discrepancy between simulation and observation at the low luminosity
end is real. The dynamical mass is $M \propto r_{1/2}\sigma^2$.  There
is good agreement between observations and simulations for $\sigma$ in
faint dwarfs. Thus, it is likely that the the reason for disagreement
in $M/L_V$ is due to the value of $r_{1/2}$, being smaller for the
observed dwarfs than the simulated ones.  This could be partly due an
observational bias that selects preferentially dSphs with higher
surface brightness and smaller $r_{1/2}$ (see
Figure~\ref{Kor}). Another explanation, is a dynamical effect not
included in our simulation that reduces the stellar radius of dwarfs
after virialization; for instance, tidal stripping or relaxation. A
larger sample size of distant dwarfs, including kinematics on And XII
and And XIV, both believed to be on their first approach to M31, would
be useful to better characterize this discrepancy.

The velocity dispersions of the stars for the observed new dwarfs and
the simulated fossils are just below $10$~km~s$^{-1}$ for luminosities
$<10^6$~L$_V$.  The scatter of the predicted and observed velocity
dispersions as a function of dwarf luminosity suggests that dwarfs
with the same luminosity may be found in a broad distribution of halos
masses.  These results are in agreement with Figure~\ref{fstar},
showing that the faintest primordial galaxies of a given luminosity
may form in halos with a total mass at virialization between
$10^6-10^7$~M$_\odot$ to a few times $10^8$~M$_{\odot}$. This is
because at these small masses the star formation efficiency is not
necessarily proportional to the dark halo mass and there is a large
scatter in $f_*$ at any given mass. This is due to the nature of
feedback effects that is local and depends on the environment (\eg,
positive feedback on the formation of H$_2$). In RG08 is found that
pre-reionization dwarfs that form in relative isolation have typically
smaller value of $f_*$ than dwarfs of the same total mass that form in
the vicinity of other luminous dwarfs (see Fig.~4 in RGS08).

\subsubsection{Understanding the luminosity-metallicity relation}

\begin{figure}
\plotone{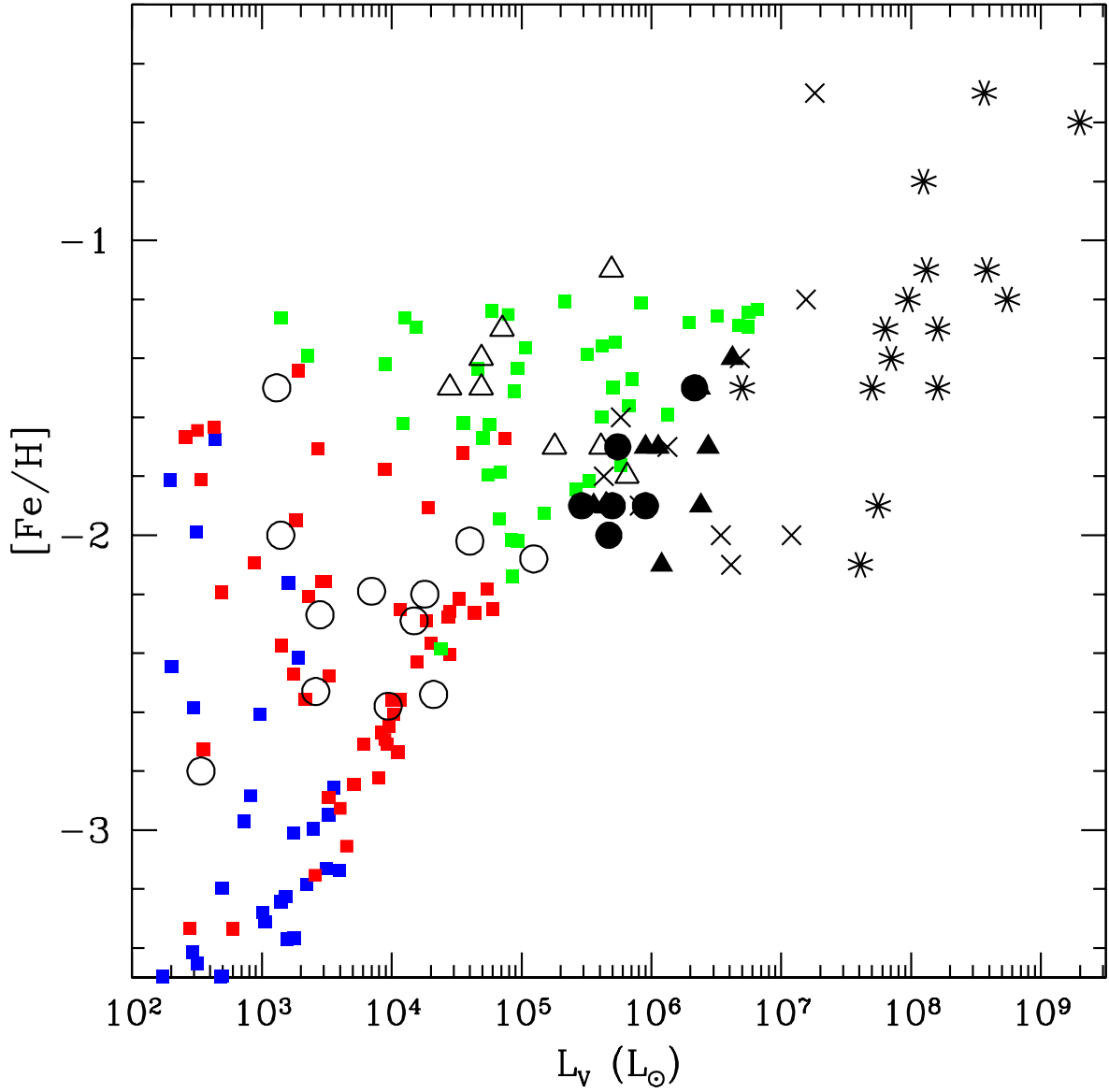}
   \caption{Metallicity vs. luminosity for the new and old
   dwarfs plotted against RG05 predictions.}
   \label{ZL}
\end{figure}

The metallicity-luminosity relation of the observed and simulated
dwarfs is shown in Figure~\ref{ZL}.  [Fe/H] is plotted against V-band
luminosity in solar units.  Symbols for the previously known dwarfs,
the new, ultra-faint dwarfs, and simulated fossils are the same as in
Figure~\ref{Kor}. In this plot, we also color code simulated fossils
according to their star formation efficiency, $f_*$, defined as
$f_*=M_*/M_{bar}$, where $M_*$ is the mass in stars and
$M_{bar}\approx M_{dm}/6$ is the baryonic mass of the halo assuming
cosmic baryons abundance. Red symbols show simulated dwarfs with
$f_*<0.003$, blue $0.003 \le f_* \le 0.03$ and green $f_*>0.03$.

The new ultra-faint dwarfs do not follow a tight
luminosity-metallicity relationship observed in more luminous galaxies
(but see \cite{Gehaetal08}). This behavior is in good agreement with
the predictions of our simulation. The recently found dwarf galaxy
Segue~1 with luminosity of $340 L_\odot$ and metallicity of [Fe/H]$\sim -2.8$
fills a gap in the luminosity-metallicity plot that was previously
devoid of observed dwarfs, present instead in the simulation \citep{Gehaetal08}. However,
Segue~1 has a half light radius that is smaller than what our
simulation predicts. Compared to their previously known counterparts,
the new dwarfs have a slightly lower metallicity but much lower
luminosities.

There are several physical mechanisms that may produce the observed
scatter in metallicities of dwarfs at a given constant luminosity.
Here, we identify the two mechanism that are dominant in our
simulation for primordial dwarf galaxies: 1) the large spread of star
formation efficiencies producing a dwarf of a given luminosity is the
dominant mechanism (to zero-th order approximation, in a closed box
model, we have $Z \propto f_*$) and 2) the existence of dwarfs
experiencing either a single or multiple episodes of star formation
contribute to the metallicity spread as well. Metal pollution from
nearby galaxies at formation might also play a role. For more massive
dwarfs that form after reionization, there may be different processes
that dominate metal enrichment. See \cite{Tassisetal08} for a
discussion.

\begin{figure}
\plotone{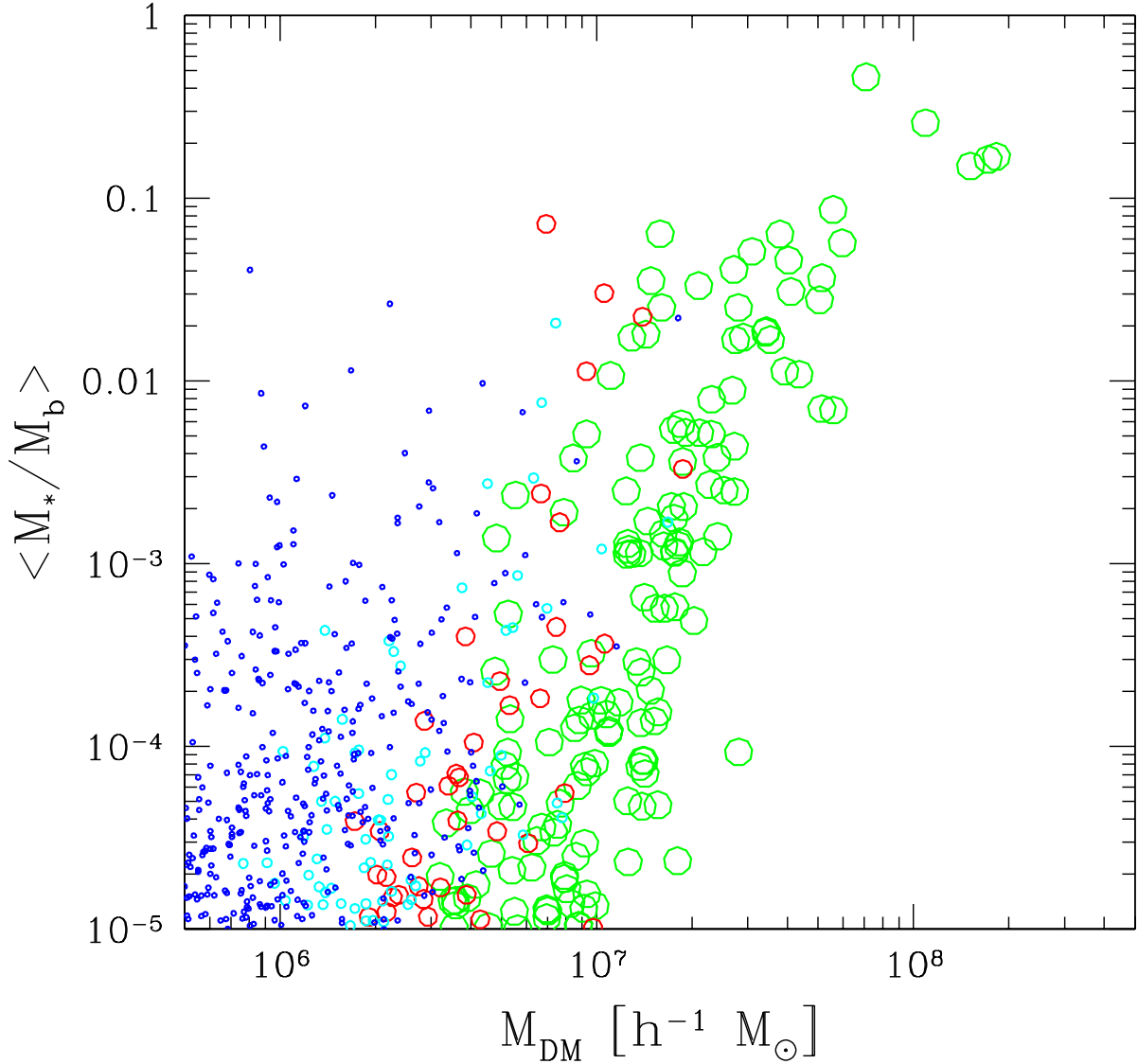}
\caption{Fraction of baryons converted into stars as function of halo
  mass of the galaxy at $z=10$.  Circles, from smaller to the larger,
  refer to galaxies with gas fractions $f_g=M_{gas}/M_b<0.1$\% (blue),
  $0.1\%<f_g<1$\% (cyan), $1\%<f_g<10$\% (red) and $f_g>10$\% (green),
  respectively.}
   \label{fstar}
\end{figure}

\begin{figure}
\plotone{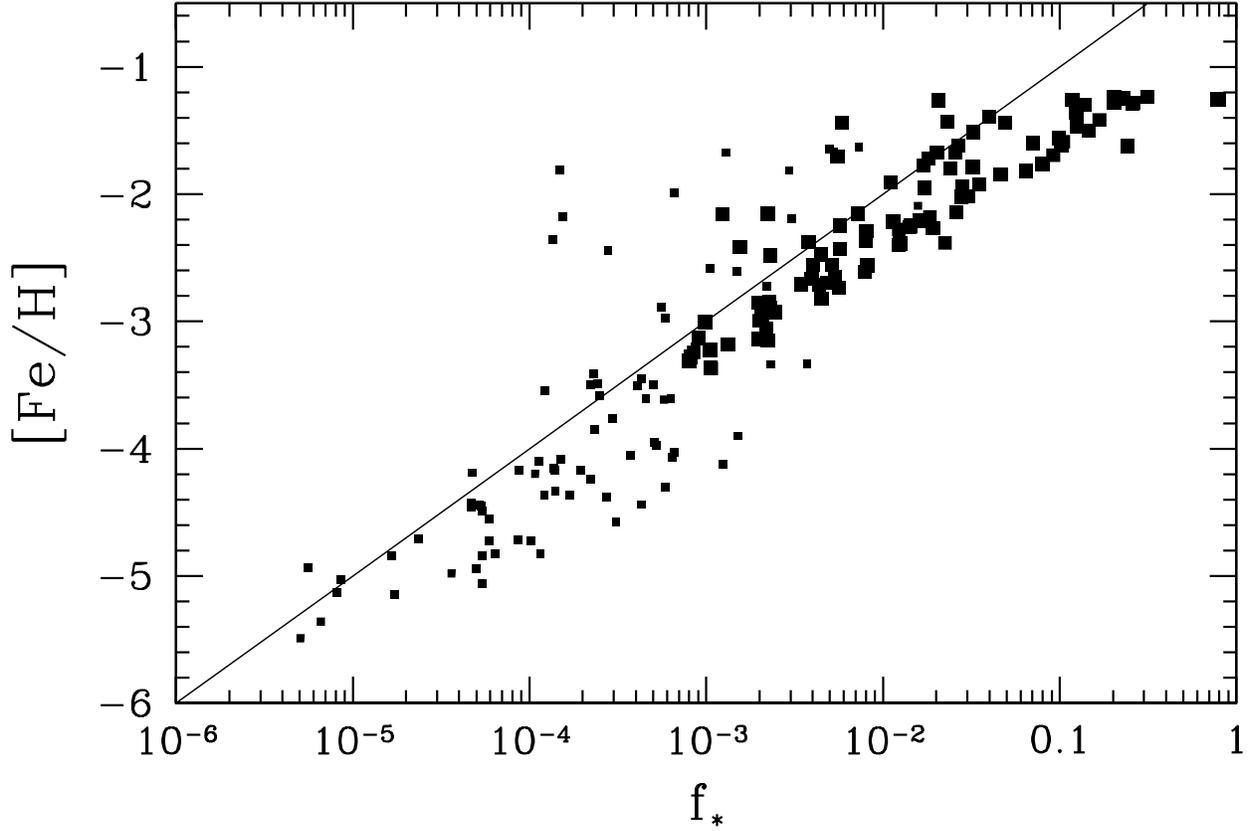}
\caption{Metallicity versus star formation efficiency $f_*=M_*/M_b$
  for the simulated fossils. The large squares show galaxies with
  $L_V\ge 10^3$~L$_\odot$, and the small squares galaxies with
  $L_V<10^3$~L$_\odot$.}
   \label{Zfs}
\end{figure}

Let's start from the first mechanism. Contrary to what it is usually
the case for more massive galaxies, we find that in primordial dwarfs
with masses $\simlt 5 \times 10^7$ M$_\odot$ the star formation
efficiency $f_*$ does not monotonically increase with halo mass (\ie,
$f_*$ has a large spread for a given halo mass or for a given mass in
stars, $M_*$, see RGS08 Figure~4 and Figure~7).  The wide range in
values results from the sensitivity of $f_*$ on a halo's environment
and is due to local feedback effects that are of fundamental
importance in determining star formation in shallow potential wells
typical of pre-reionization dwarfs.  Figure~\ref{fstar} (taken from
Figure~7 in RGS08) shows that below a few
$10^7$~M$_\odot$, halos with exactly the same dark mass can be either dark or luminous (depending on the
environment). Thus, feedback
from non-ionizing and ionizing UV radiation, mechanical feedback and
chemical enrichment can produce two halos with the same dark mass and
very different star formation efficiencies.  This appears to be the
main effect responsible for the observed spread in metallicity for a
given luminosity, in the new dwarfs.  In Figure~\ref{Zfs}, we plot the
metallicity as a function of the mean star formation efficiency for
the halo, $f_*$. As expected, simulated dwarfs with higher values of
the star metallicity are the ones with the larger value of $f_*$.

However, this effect alone cannot account for all the observed scatter
of the metallicity as illustrated by the color coding of simulated
dwarfs in Figure~\ref{ZL}. It appears that the metallicity is not
simply proportional to $f_*$ (otherwise the boundaries between symbols
of different color would be horizontal).  Instead, for a given value of
$f_*$, the metallicity is larger for fainter dwarfs.  It is not too
surprising that $Z$ is not simply proportional to $f_*$. Even when
using a very simple chemical evolution model, neglecting gas inflows
and assuming instantaneous metal recycling, the mean metallicity of
the stars is proportional to $M_*/M_{gas}$ rather than
$f_*=M_*/M_{bar}$, where $M_{gas}$ is the initial value of the gas
mass available for star formation. Thus, $Z \propto f_*
(M_{bar}/M_{gas})$. If feedback effects reduce the value of
$M_{gas}/M_{bar}$ below unity in the smallest and lowest luminosity
primordial dwarfs, the metallicity of the stars will be larger for a
fixed value of $f_*$, as observed in Figure~\ref{Zfs} and
Figure~\ref{ZL}. The reduction of $M_{gas}/M_{bar}$ below unity can be
produced by three effects: the increase of the Jeans mass of the IGM
over the virial mass of the halo due to reheating (see Figure~6 in
RGS08), heating of the gas via ionizing radiation from stars within
the halo, and by multiple episodes of star formation with a first
burst that lowers $M_{gas}$ substantially, but does not produce
sufficiently large values of $f_*$ and $Z$ when compared to subsequent
bursts.

\begin{figure}
\plotone{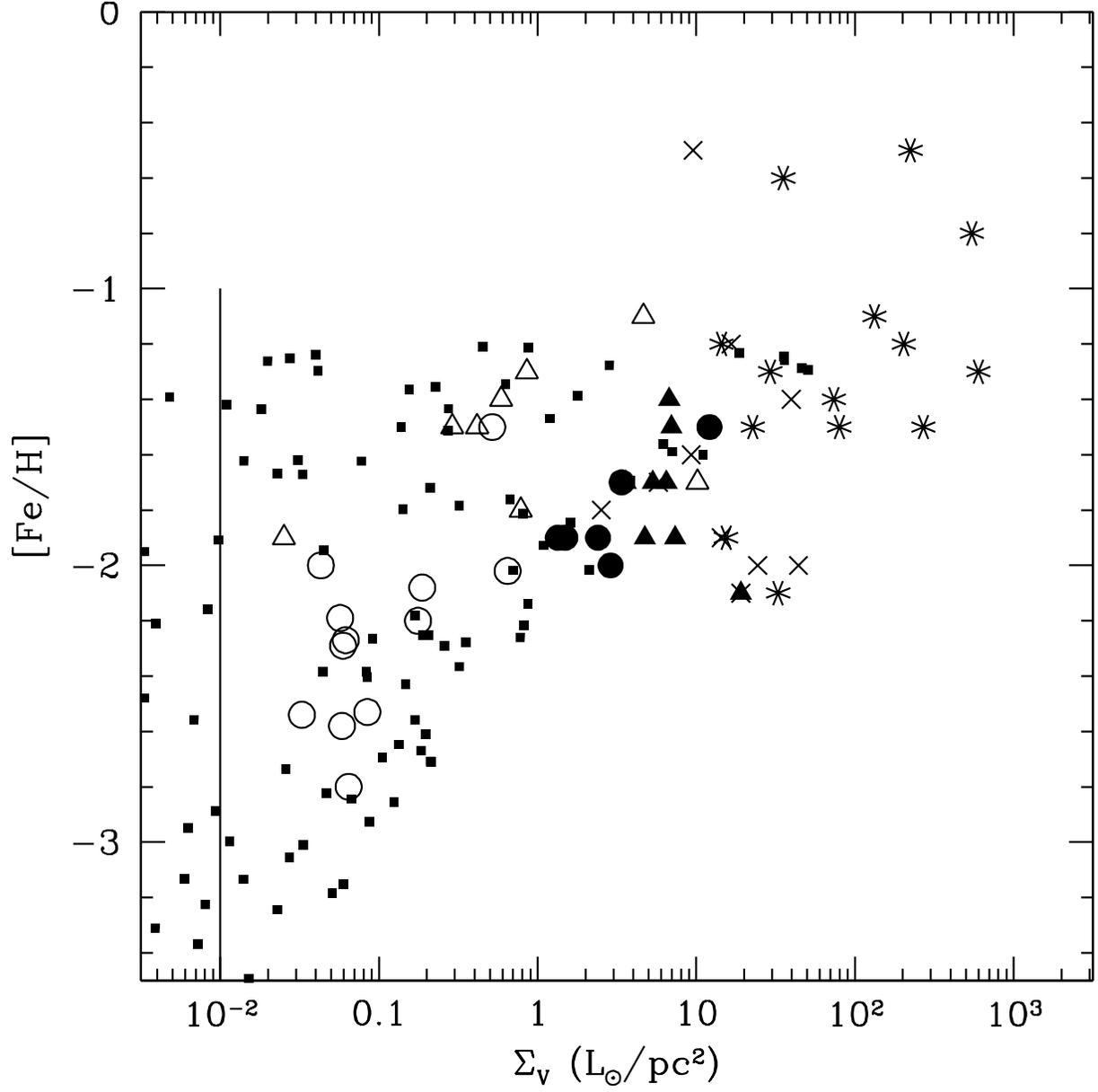}
   \caption{Metallicity vs. surface brightness for the new and old
     dwarfs.  The surface brightness limit of the SDSS is shown by a
     solid vertical line. Note the predicted dwarfs with Z$<$-2.5 and surface
   brightnesses above Sloan detection limits.}
   \label{ZS}
\end{figure}

Figure~\ref{ZS} shows [Fe/H] versus the surface brightness in the
V-band, $\Sigma_V$. The symbols are the same as in Fig.~\ref{Kor} and
the solid line shows the SDSS sensitivity limit. No trend is observed
between metallicity and $\Sigma_V$ for the simulated dwarfs. Observed
dwarfs show less scatter for $\Sigma_V$-metallicity relation than for
the luminosity-metallicity relation in Figure~\ref{ZL}. There is one
dwarf with metallicity below [Fe/H]$=-2.5$: Segue-1 that
has [Fe/H]$=-2.8$. Since the spectral synthesis method used in
\cite{Kirby:08} and \cite{Gehaetal08} may not be subject to the
overestimation of metallicities seen with measurements using the CA
triplet, the lack of dwarfs with [Fe/H]$< -3.0$ could be a sign of a
change in the IMF at very low [Fe/H].

Finally, we have examined whether there is a dependence of the
metallicities on the distance of the galaxy from the Milky Way or
Andromeda.  For the new Milky Way dwarfs there is slight trend of
higher metallicities at smaller Galactocentric distances; however, the
upward trend is dominated by Ursa Major II and Coma Ber., both of of
which show evidence for tidal disruption.  For the M31 dwarfs, no
trend was observed.

\begin{figure}
\plotone{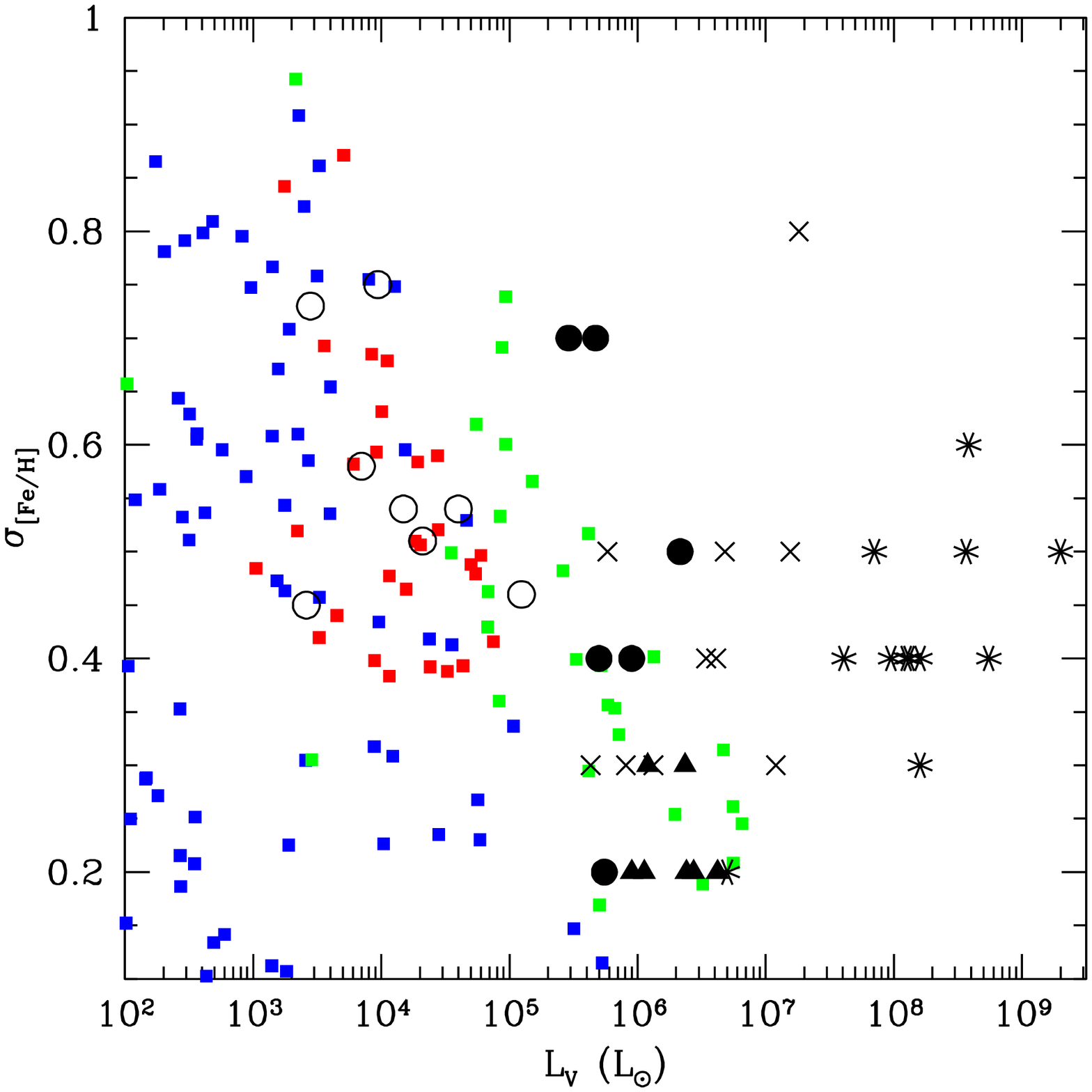}
\caption{Metallicity spread vs. V-band luminosity for the new and old
  dwarfs. Simulation data for the metallicity spread may be
  unrelayable for dwarfs with luminosities $L_V <
  10^3-10^{4}$~L$_\odot$, due to the small number of stellar particles
  in the galaxies.}
   \label{sZL}
\end{figure}

Figures~\ref{sZL} show the scatter of the metallicity of the stars,
$\sigma_{[Fe/H]}$, plotted against V-band luminosity and [Fe/H]
respectively.  Once more, the various point types and colors are the
same used in Figure~\ref{ZL}.  The observational data for the new
dwarfs matches the predictions, though Figure~\ref{sZL} shows a lack
of low $L_V$ objects with $\sigma_{[Fe/H]} < 0.4$.  However, given the
small number of data points available, it is not possible to rule out
selections effects of small number statistics as an explanation.
Dwarfs with low values of $\sigma_{[Fe/H]}$ tend to have higher
luminosities and are equally likely to be a dE or dSphs.  However,
dwarfs with the highest $\sigma_{[Fe/H]}$ are faint dSphs with $L_V <
10^{6}$~L$_\odot$, and occupy the lowest mass dark matter halos. It is
not clear at this point how reliable the simulation data for
$\sigma_{[Fe/H]}$ is for dwarfs with luminosities $L_V <
10^3-10^{4}$~L$_\odot$. However, all the dwarfs we analyze have at
least 10 stellar particles and 100 dark matter particles. The masses
of stellar particles vary, depending on the star formation efficiency
and the duration of the star burst.

As with the metallicities, we looked at how the metallicity spread
depends on the distance from the host.  For both previously known and
ultra-faint dSphs there is no dependence on distance within the virial
radius of the Milky Way.  There is a lack of dwarfs with high
$\sigma_{[Fe/H]}$ beyond 400~kpc; however, given the small number of
data points and the luminosity and surface brightness limits of
current surveys, the trend is not statistically significant.


\subsection{The Missing Galactic Satellite Problem Revisited}

\begin{figure}
\plotone{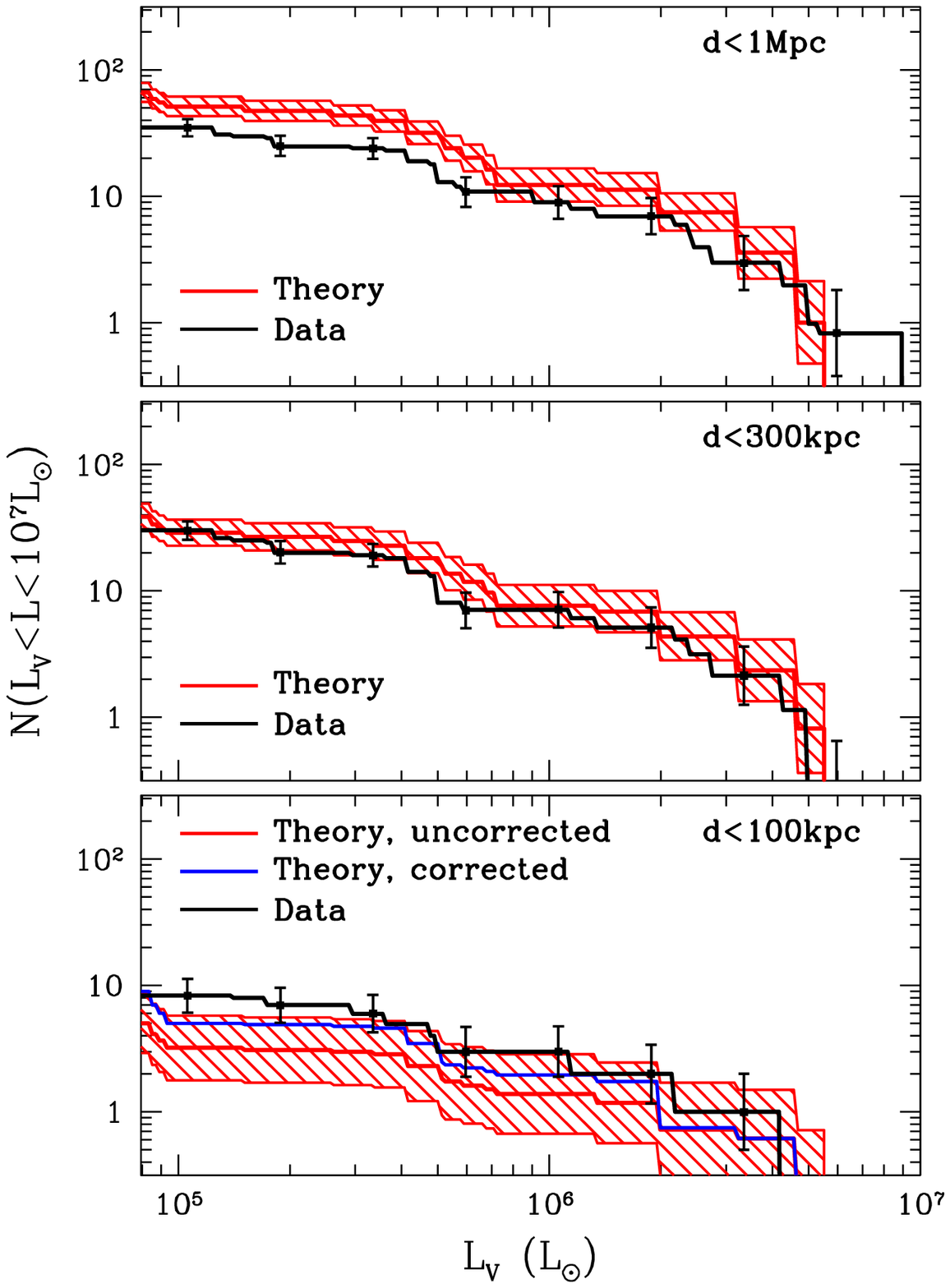}
\caption{Luminosity function of pre-reionization fossil dwarfs
  predicted in GK06 (red) plotted with the luminosity function for the
  new and old Local Group dSphs.  The black lines are the
  observations corrected for completeness as discussed in Section 2.
  Corrections to the theory account for an under-abundance of small
  halos near the hosts due to numerical effects (GK06).}
   \label{LF}
\end{figure}

In order to simulate a representative sample of the universe, the
size of cosmological simulations must be significantly larger than the
largest scale that becomes non-linear at the redshift of interest.  At
$z = 0$, this scale is at least $50$ to $100$~Mpc.  Current
computational resources are not able to evolve a cosmological
simulation of this size that includes all the relevant gas and
radiation physics, to $z = 0$.  However, as argued in \S~3, the
properties of those fossil galaxies that survive tidal destruction
change only through passive aging of their stars formed before
reionization (but see Ricotti 2008), allowing their
properties at $z = 0$ to be simply related to their properties at
reionization.  GK06 uses this approximation in conjunction with
high-resolution N-body simulations of the Local Group, to evolve a
population of dwarf galaxies around a Milky Way mass halo from $z =
70$ to $z = 0$.  For details of the simulations, see \S~2 in GK06.

GK06 define a fossil as a simulated halo which survives to $z = 0$
and remains below the critical circular velocity of $20$~km~s$^{-1}$
with no appreciable tidal stripping.  They calculate the probability,
$P_S(v_{c, max}, r)$, of a luminous halo with a given maximum circular
velocity $v_{c, max}$ to survive from $z = 8$ (the final redshift of the
RG05 simulation) to $z = 0$.  For a given $v_{c, max}$, the number of
dwarfs at $z = 0$ is $N(v_{c, max}, z = 8)P_S(v_{max},r)$. The surviving
halos are assigned a luminosity based on the $L_V$ versus $v_{c, max}$
relationship from RG05.  At $z = 0$, GK06 has a population of dwarf
galaxies with a resolution limit of $v_{c, max} = 13$~km~s$^{-1}$.
Unfortunately, this limit corresponds to a lower luminosity limit of
$L_V \sim 10^5$~L$_{\odot}$, which includes Leo T and Canes Venatici
I, but excludes all the other new ultra-faint Milky Way satellites. We
do not show the prediction of the GK06 model below this lower
luminosity limit. A new N-body method with a resolution limit of $L_V
\sim 10^3$~L$_{\odot}$ that follows the merger histories for the halos
will be presented in Bovill \& Ricotti (2008), in preparation.

In Figure~\ref{LF}, we show the cumulative luminosity function from
GK06 for the Milky Way and M31 satellites with the addition of the new
ultra-faint dSphs.  The lower panel shows satellites with distance
from their host $d<100$~kpc, the middle panel $d<300$~kpc and the
upper panel $d<1$~Mpc.  The gray lines show the GK06 predictions, and
the shaded region encompasses the error bars.  Since, the resolution
limits of GK06 causes halos with $v_{c, max}<17$~km~s$^{-1}$ to the
preferentially destroyed by tidal effects, the predicted luminosity
function is corrected.  Both the uncorrected (lower) and corrected
(upper) luminosity functions are plotted in the lower panel.  On all
three panels, the black histogram and the points with error bars
represent the observed luminosity function of fossil dwarfs around the
Milky Way and M31.  Their numbers have been corrected for completeness
as discussed in \S~2.  For the purposes of this plot, we are
considering all the new dwarfs to be fossil candidates.

For $d <100$~kpc (bottom panel), there is an overabundance of observed
satellites with respect to the simulated luminosity function from
$10^5$ to $10^6$~L$_{\odot}$. This discrepancy is likely due to
excessive destruction rate of satellites caused by the insufficient
resolution the GK06 N-body simulations. At distances $d <300$~kpc
(middle panel), there is excellent agreement between theory and
observation.  Canes Venatici I and the new Andromeda satellites are
included in the latter panel.  The upper panel shows the luminosity
function for all dwarfs within $1$~Mpc of the host, including Leo T.
Note, that GK06 assumes an isolated Milky Way type galaxy (with total
mass comparable to the Local Group mass), while observations with
$d<1$~Mpc of the Milky Way include the satellite system around M31.
For $d < 1$~Mpc, there is an under-abundance of observed satellites
between $10^5$ to $10^6$~L$_{\odot}$ with respect to the simulation
predictions.  However, this is consistent with the theory since beyond
$250$~kpc dwarfs with $L_V \sim 10^5 L_{\odot}$ drop below SDSS
detection limits (Koposov et al, 2007). Hence, the under-abundance of
observed dwarfs at large distances may be due to the completeness
limit of the survey.

\section{Discussion and Conclusions}\label{sec:disc}

There are two main ideas for the origin of dSphs in the Local
Group. Most importantly, these two ideas have very different
implications for models of galaxy formation, and the minimum mass a
dark halo needs to host a luminous galaxy. The ``tidal scenario'',
holds the dwarfs we see today were once far more massive, having been
stripped of most of their dark matter during interactions with larger
galaxies \citep[\eg,][]{KravtsovGnedinKlypin04}.  In this model, we
would expect the halos with original dark matter masses below $10^{8}$
~M$_{\odot}$ to be mostly dark at formation and at the modern epoch.
The ``primordial scenario'', has dwarf galaxies starting with close to
their current stellar mass of about $10^3-10^6$~M$_{\odot}$ and, with
several dark halos with mass at formation below the threshold of about
$2 \times 10^{8}$~M$_{\odot}$ hosting a luminous galaxy.  Star
formation in halos this small is possible only before reionization and
is widespread if ``positive'' feedback plays a significant role in
regulating star formation in the first galaxies
\citep{RicottiGnedinShull01,RicottiGnedinShull02a,RicottiGnedinShull02b}.

In this paper, we argue that the recent discovery of the ultra-faint
dwarfs in the Milky Way and M31 supports the ``primordial
scenario''. The existence of the ultra-faint dwarfs was predicted by
simulations of the formation of the first galaxies (see RG05) and, as
shown in the present work, the observed properties of this new
population are consistent with them being the ``fossils'' of the first
galaxies.

While tidal stripping can reproduce properties of an individual
galaxy, it is unable to completely reproduce all the trends in the
ultra-faint population.  This is primarily seen in the kinematics of
the ultra-faint dwarfs.  Tidal stripping predicts a steeper than
observed drop in $\sigma$ with $L_V$ \citep{PenarrubiaNM08}, while our
simulations show primordial dwarfs which match the observed trends in
$\sigma$ extremely well. It has not been shown yet that star formation
in dwarf galaxies more massive than $10^8-10^9$~M$_\odot$ can
reproduce the observed properties of ultra-faint dwarfs without requiring
tidal stripping of stars.

The tidal model predicts that gas rich dIrr loose their gas and
transform into dSphs due to tidal or ram pressure interaction with a
host halo.  And XII, which shows a proper motion close to current
published escape velocity of M31, may be on its first approach to the
Local Group \citep{Martinetal06, Chapmanetal07}.  A similar situation
exists for And XIV.  With a dynamical mass of $M \sim 3\times10^7
M_{\odot}$, And XIV has $v > v_{esc}^{M31}$, suggesting it is also
just entering the Local Group \citep{Chapmanetal07}.  In the tidal
model \citep{Mayeretal07,Mayeretal06}, And XII and And XIV would be
expected to still harbor significant reservoirs of gas, however,
observations show And XIV has $M_{HI} < 3\times10^3 M_{\odot}$
\citep{Chapmanetal07} and And XII has no detected \HI~
\citep{Martinetal06}.  If neither of these dwarfs have undergone
significant tidal interactions with their hosts, as their velocities
suggest, how did they lose their gas?  Though its velocity is unknown,
the recently discovered And XVIII \citep{McConnachieetal08}, shows the
same characteristics.  At a distance of $600$~kpc from M31 and 1.35
Mpc from the Milky Way, it is unlikely that And XVIII has undergone
significant interaction with either Local Group spirals.  And XVIII is
classified as a dSph with no detected \HI~and is similar to the Cetus
and Tucana dwarfs \citep{McConnachieetal08}, both of which are good
candidate fossil galaxies (RG05).

On the opposite end of the spectrum is the strange case of Leo~T, the
properties of which are discussed in Section 2.1.1.  While, Leo~T has
an \HI~mass fraction typical of dIrr, its other properties are
indistinguishable from the other newly discovered ultra-faint dwarfs
\citep{SimonGeha07}, all of which are dSph.  Leo~T's large distance
from its host, \HI~reservoir and low probability of recent tidal
interactions \citep{deJongetal08} make it a good candidate for a
precursor to a dSph in the tidal scenario.  Particularly given that
Leo T's dynamical mass within the stellar spheroid is small:
$8.2\times10^6$~M$_\odot$ \citep{SimonGeha07}, its gas is unlikely to
survive a single tidal encounter intact.  Therefore, Leo T may have
formed at or near its current mass, and the striking similarity of Leo
T to its ultra-faint counterparts suggests that they too could have
formed as primordial dwarfs at their current masses.
 
By our definition, pre-reionization fossils are dwarfs that form
before reionization in dark halos with $v_{c}<v_c^{cr}\sim
20$~km~s$^{-1}$, while non-fossils dwarfs form in halos with
$v_{c}>v_c^{cr}$ before and after reionization. The value
$v_c^{cr}\sim 20$~km~s$^{-1}$ that we use to define a fossil is
primarily motivated by fundamental differences in cooling and feedback
processes that regulate star formation in these halos in the early
Universe. This value of the circular velocity is also very close to
estimates based on the suppression of star formation in dwarfs after
reionization \citep{Gnedin-filteringmass00, OkamotoGT08}. However, as
argued in \citep{Ricotti:08}, fossils dwarfs can have a late phase of
gas accretion and star formation well after reionization, at redshift
$z<1-2$. Thus, a complete suppression of star formation after
reionization is not necessarily what defines a ``fossil galaxy''.

The number of Milky Way dark satellites that have or had in the past
$v_{c}>v_c^{cr}$ can be estimated using the results of published
N-body simulations (see \S~\ref{ssec:count}). We find that using the
via Lactea N-body simulation there are approximately $N_{dark} \approx
73 \pm 16$ halos with $v_{c}>20$~km~s$^{-1}$ within the virial radius
\citep{Diemandetal07a}.  The new Aquarius simulations
\citep{Springeletal08}, however, show a factor of 2.5 increase in the
number of halos with $v_{c}>20$~km~s$^{-1}$, \ie, $N_{dark}\sim 182
\pm 40$ dark halos. Within a distance of $200$~kpc we estimate
$N_{dark} \approx 36 \pm 8$ for the Via Lactea and $N_{dark} \approx
91 \pm 20$ for the Aquarius simulation.

If the number of observed dwarf satellites within the Milky Way (after
applying completeness corrections) is larger than $N_{dark}$ we must
conclude that some satellites are fossils.  Twelve new ultra-faint
dwarfs have been discovered around the Milky Way by analyzing SDSS
data in a region that covers about 1/5 of the sky. Applying a simple
correction for the sky coverage we estimate that there should be about
at least $85 \pm 14$ Milky Way satellites. However, the data becomes
incomplete for ultra-faint dwarf that are further than about $200$~kpc
from the Galactic center. Comparing this number of luminous satellites
to $N_{dark}$ within $200$~kpc we cannot conclusively conclude that
some ultra-faint dwarfs are fossils because $N_{dark}$ for the
Aquarius simulation is comparable to the estimate number of luminous
satellites.

Once both sensitivity and survey area corrections are applied,
\cite{Tollerudetal08} estimates the existence of $300$ to $600$
luminous satellites within the virial radius ($R_{vir} \sim 400$ kpc)
of the Milky Way and $120$ within $200$~kpc. Comparing $N_{dark}$ to
the \cite{Tollerudetal08} estimates of the number of luminous Milky
Way satellites implies that a significant fraction of them are fossils
(regardless if we use the Via Lactea or the Aquarius simulations
estimates for $N_{dark}$).  In Table~\ref{tab:count} we have
summarized the aforementioned results.

Another argument for the existence of fossils is provided by detailed
comparison of the Galactocentric distribution of fossils in the Milky
Way (GK06). Based on these comparison GK06 find that about $1/3$ of
Milky Way dwarfs may be fossils.  In this paper, we show the GK06
theoretical results in comparison to updated observational data,
including the new ultra-faint dwarfs found using SDSS data, and
applying completeness correction due to the limited area surveyed by
the SDSS (about 1/5 of the sky).  Assuming that the Local Group has a
mass of $3 \times 10^{12}$ M$_{\odot}$, as in the GK06 simulation, we
find that there are no ``missing galactic satellites'' with $L_V \geq
10^5$~L$_{\odot}$ within the virial radius of the Milky Way.  When the
new dwarfs are included, the observed and predicted numbers of
satellites agree near the Milky Way, however, for distances greater
than $200$~kpc, it is clear that there is still a 'missing' population
of dwarfs.  However, given that for $d >200$~kpc, dwarfs with $L_V
\sim 10^5$~L$_{\odot}$ drop below SDSS detection limits (Koposov et
al, 2007), the under-abundance of observed dwarfs at large distances
is not surprising and likely due to the SDSS sensitivity limit.

A final comment regards the cosmological model. The RG05 and GK06
simulations use cosmological parameters from WMAP~1.  N-body
simulations show that the number $N(M)$ of Milky Way dark matter
satellites as a function of their mass is not overly sensitive to the
cosmology, although there are some differences on the number of the
most massive satellites \citep{Madauetal08}.  However, $N(v_{max})$
should be sensitive to the cosmology \citep{ZentnerBullock03}, and
changes of $\sigma_8$ and $n_s$ may affect the occupation number and
Galactocentric distribution of luminous halos.  The collapse time of
small mass halos in high density regions probably dominates the 20\%
variations in $\sigma_8$ between WMAP~1 and WMAP~3, limiting effects
due to the cosmology near large halos.  A decrease in luminous dwarf
numbers, due to the lower $\sigma_8$, could be evident in the
distribution of the lowest mass luminous halos in the voids.

In conclusion, the number of Milky Way and M31 satellites provides an
indirect test of galaxy formation and the importance of positive
feedback in the early universe. Although the agreement of the SDSS and
new M31 dwarfs' properties with predictions from the RG05 and GK06
simulations does not prove the primordial origin of the new
ultra-faint dwarfs, it supports this possibility with quantitative
data and more successfully than any other proposed model has been able
to do so far. At the moment, we do not have an ultimate
observational test that can prove a dwarf galaxy to be a fossil. Even
a test based on measuring the SFH of the dwarf galaxies may not be
discriminatory because, as has been recently suggested, fossil
galaxies may have a late phase of gas accretion and star formation at
$z<1-2$, during the last $9-10$~Gyrs \citep{Ricotti:08}. The
distinction between fossils and non-fossils galaxies thus is quite
tenuous and linked to our poor understanding of star formation
and feedback in dwarf galaxies. Arguments based on counting the number
of dwarfs in the Local Universe probably provide the most solid
argument to prove or disprove the existence of fossil galaxies. In the
future, a possible test may be provided by deep surveys looking for
ultra-faint or dark galaxies in the local voids. Some fossil dwarfs
should be present in the voids if they formed in large numbers before
reionization.

\bibliographystyle{../../apj}
\bibliography{./2ndyear}

\begin{deluxetable}{cccccc}
\tablecaption{Summary of New Dwarfs Properties}\label{tab:one}
\tablewidth{0pt}
\tablecolumns{6}
\tablehead{
\colhead{Dwarf} &
\colhead{Host} &
\colhead{$d_{hel}$ ($d_{M31}$)} &
\colhead{$M_{V}$} &
\colhead{$\mu_{o}$} & 
\colhead{$r_{1/2}$} \\

\colhead{} &
\colhead{} &
\colhead{($kpc$)} & 
\colhead{($mag$)} &
\colhead{($mag$ $arcsec^{-2}$)} & 
\colhead{($pc$)}

}
\startdata
Bootes I & MW & 62 $\pm$ 3 \tablenotemark{\rm a} & -5.8 $\pm$ 0.5 \tablenotemark{\rm f} & 28.3 $\pm$ 0.5 \tablenotemark{\rm f} & 230 \tablenotemark{\rm a} \\
Bootes II & MW & 60 $\pm$ 10 \tablenotemark{\rm b} & -3.1 $\pm$ 1.1 \tablenotemark{\rm b} & 29.8 $\pm$ 0.8 \tablenotemark{\rm b} & 72 $\pm$ 28 \tablenotemark{\rm b} \\
Canes Venatici I & MW & 220$^{+25}_{-16}$ \tablenotemark{\rm c} & -7.9 $\pm$ 0.5 \tablenotemark{\rm c} & 28.2 $\pm$ 0.5 \tablenotemark{\rm c} & 550 \tablenotemark{\rm c} \\
Canes Venatici II & MW & 150$^{+15}_{-14}$ \tablenotemark{\rm d} & -4.8 $\pm$ 0.6 \tablenotemark{\rm d}& - & $\sim$140 \tablenotemark{\rm d} \\
Coma Berenics & MW & 44 $\pm$ 4 \tablenotemark{\rm d} & -3.7 $\pm$ 0.6 \tablenotemark{\rm d} & - & 70 \tablenotemark{\rm d} \\
Hercules & MW & 140$^{+13}_{-12}$ \tablenotemark{\rm d} & -6.0 $\pm$ 0.6 \tablenotemark{\rm d} & - & $\sim$320 \tablenotemark{\rm d} \\
Leo IV & MW & 160$^{+15}_{-14}$ \tablenotemark{\rm d} & -5.1 $\pm$ 0.6 \tablenotemark{\rm d} & - & $\sim$160 \tablenotemark{\rm d} \\
Leo T & MW & $\sim$420 \tablenotemark{\rm e} & -7.1 \tablenotemark{\rm e} & 26.9 \tablenotemark{\rm e} & $\sim$170 \tablenotemark{\rm e} \\
Ursa Major I & MW & 106$^{+9}_{-8}$ \tablenotemark{\rm g} & -5.6 $\pm$ 0.6 \tablenotemark{\rm g} & - & 308 $\pm$ 32 \tablenotemark{\rm g} \\
Ursa Major II & MW & 30 $\pm$ 5 \tablenotemark{\rm a}& -3.8 $\pm$ 0.6 \tablenotemark{\rm h} & - & 127 $\pm$ 21 \tablenotemark{\rm g} \\
Segue I & MW & 23 $\pm$ 2 \tablenotemark{\rm n} & -1.5$^{+0.6}_{-0.8}$ \tablenotemark{\rm n} & -
& 29 $\pm$ 8 \tablenotemark{\rm n} \\
Willman I & MW & 38 $\pm$ 7 \tablenotemark{\rm i}& -2.5 $\pm$ 1.0 \tablenotemark{\rm i} & - & 21 $\pm$ 5 \tablenotemark{\rm i} \\
\\
And XI & M31 & $\sim$770 \tablenotemark{\rm j} & -7.3 $\pm$ 0.5 \tablenotemark{\rm j} & - & 115 $\pm$ 45 \tablenotemark{\rm j} \\
And XII & M31 & 830 $\pm$ 50 ($\sim$105) \tablenotemark{\rm k} & -6.9 \tablenotemark{\rm k} & - & 137 \tablenotemark{\rm k} \\
And XIII & M31 & $\sim$770 \tablenotemark{\rm j} & -6.9 $\pm$ 1.0 \tablenotemark{\rm j} & - & 115 $\pm$ 45 \tablenotemark{\rm j} \\
And XIV & M31 & 871 \tablenotemark{\rm m}  & - & - & - \\
And XV & M31 & 630 $\pm$ 60 (170) \tablenotemark{\rm l} & -9.4 \tablenotemark{\rm l} & - & - \\
And XVI & M31 & 525 $\pm$ 50 (270) \tablenotemark{\rm l}  & -9.2 \tablenotemark{\rm l} & - & - \\
\enddata

\tablenotetext{a}{\cite{Martinetal07}}
\tablenotetext{b}{\cite{Walshetal07}}
\tablenotetext{c}{\cite{Zuckeretal06a}}
\tablenotetext{d}{\cite{Belokurovetal07}}
\tablenotetext{e}{\cite{Irwinetal07}}
\tablenotetext{f}{\cite{Belokurovetal06a}}
\tablenotetext{g}{\cite{SimonGeha07}}
\tablenotetext{h}{\cite{Zuckeretal06b}}
\tablenotetext{i}{\cite{Willmanetal05AJ}}
\tablenotetext{j}{\cite{Martinetal06}}
\tablenotetext{k}{\cite{Chapmanetal07}}
\tablenotetext{l}{\cite{Ibataetal07}}
\tablenotetext{m}{\cite{Majewskietal07}}
\tablenotetext{n}{\cite{Gehaetal08}}

\end{deluxetable}

\begin{deluxetable}{ccccccc}
\tablecaption{Summary of New Dwarfs Properties cont.}\label{tab:two}
\tablewidth{0pt}
\tablecolumns{7}
\tablehead{
\colhead{Dwarf} &
\colhead{Host} &
\colhead{Type} &
\colhead{[Fe/H]} &
\colhead{$\sigma$} &
\colhead{$M_{TOT}$} & 
\colhead{M/L} \\
\colhead{} &
\colhead{} &
\colhead{} &
\colhead{} & 
\colhead{($km s^{-1}$)} &
\colhead{($10^{6} M_{\odot}$)} & 
\colhead{}
}
\startdata
Bootes I & MW & dSph & -2.1 \tablenotemark{\rm a} & 6.5$_{-1.4}^{+2.0}$ \tablenotemark{\rm a} & 13 \tablenotemark{\rm a} & $\sim$700 \tablenotemark{\rm a} \\
Bootes II & MW & dSph & -2.0 \tablenotemark{\rm c} & - & - & - \\
Canes Venatici I & MW & dSph & -2.09 $\pm$ 0.02 \tablenotemark{\rm b} & 7.6 $\pm$ 0.4 \tablenotemark{\rm b} & 27 $\pm$ 4 \tablenotemark{\rm b} & 221 $\pm$ 108 \tablenotemark {\rm b} \\
Canes Venatici II & MW & dSph & -2.31 $\pm$ 0.12 \tablenotemark{\rm b} & 4.6 $\pm$ 1.0 \tablenotemark{\rm b} & 2.4 $\pm$ 1.1 \tablenotemark{\rm b} & 336 $\pm$ 240 \tablenotemark{\rm b} \\
Coma Berenics & MW & dSph & -2.0 $\pm$ 0.07 \tablenotemark{\rm b} & 4.6 $\pm$ 0.8 \tablenotemark{\rm b} & 1.2 $\pm$ 0.4 \tablenotemark{\rm b} & 448 $\pm$ 297 \tablenotemark{\rm b} \\
Hercules & MW & dSph &  -2.27 $\pm$ 0.07 \tablenotemark{\rm b} & 5.1 $\pm$ 0.9 \tablenotemark{\rm b} & 7.1 $\pm$ 2.6 \tablenotemark{\rm b} & 332 $\pm$ 221 \tablenotemark{\rm b} \\
Leo IV & MW & dSph & -2.31 $\pm$ 0.1 \tablenotemark{\rm b} & 3.3 $\pm$ 1.7 \tablenotemark{\rm b} & 1.4 $\pm$ 1.5 \tablenotemark{\rm b} & 151 $\pm$ 177 \tablenotemark{\rm b} \\
Leo T & MW & dSph/dIrr & -2.29 $\pm$ 0.1 \tablenotemark{\rm b} & 7.5 $\pm$ 1.6 \tablenotemark{\rm b} & 8.2 $\pm$ 3.6 \tablenotemark{\rm b} & 138 $\pm$ 71 \tablenotemark{\rm b} \\
Ursa Major I & MW & dSph & -2.06 $\pm$ 0.1 \tablenotemark{\rm b} & 7.6 $\pm$ 1.0 \tablenotemark{\rm b} & 15 $\pm$ 4 \tablenotemark{\rm b} & 1024 $\pm$ 636 \tablenotemark{\rm b} \\
Ursa Major II & MW & - & -1.97 $\pm$ 0.15 \tablenotemark{\rm b} & 6.7
$\pm$ 1.4 \tablenotemark{\rm b} & 4.9 $\pm$ 2.2 \tablenotemark{\rm b}
& 1722 $\pm$ 1226 \tablenotemark{\rm b} \\
Segue I & MW & dSph & -2.8 $\pm$ 0.2 \tablenotemark{g} & 4.3 $\pm$ 1.2
\tablenotemark{g} & 0.45$^{+4.7}_{-2.5}$ \tablenotemark{g} &
1320$^{+2680}_{-940}$ \tablenotemark{g} \\
Willman I & MW & dSph & -1.5 \tablenotemark{\rm a} & 4.3$_{-1.3}^{+2.3}$ \tablenotemark{\rm a} & 0.5 \tablenotemark{\rm a} & $\sim$470 \tablenotemark{\rm a} \\
\\
And XI & M31 & - & -1.3 $\pm$ 0.5 \tablenotemark{\rm d} & - & - & - \\
And XII & M31 & dSph & -1.5 $\pm$ 0.4 \tablenotemark{\rm d} & - & - & - \\
And XIII & M31 & - & -1.4 $\pm$ 0.5 \tablenotemark{\rm d} & - & - & - \\
And XIV & M31 & dSph & -1.7 \tablenotemark{\rm e} & 6.2 $\pm$ 1.3 \tablenotemark{\rm e} & 33$^{+31}_{-8}$ \tablenotemark{\rm e} & 165$^{+156}_{-92}$ \tablenotemark{\rm e} \\
And XV & M31 & - & -1.1 \tablenotemark{\rm f} & - & - & - \\
And XVI & M31 & - & -1.7\tablenotemark{\rm f}  & - & - & - \\
\enddata

\tablenotetext{a}{\cite{Martinetal07}}
\tablenotetext{b}{\cite{SimonGeha07}}
\tablenotetext{c}{\cite{Walshetal07}}
\tablenotetext{d}{\cite{Martinetal06}}
\tablenotetext{e}{\cite{Majewskietal07}}
\tablenotetext{f}{\cite{Ibataetal07}}
\tablenotetext{g}{\cite{Gehaetal08}}

\end{deluxetable}

\end{document}